\documentclass[apj]{emulateapj}

\usepackage{longtable,lscape}
\usepackage{rotating}
\usepackage{color}
\definecolor{orange}{rgb}{1,0.5,0}
\definecolor{purple}{rgb}{0.5,0.1,0.7} 
\definecolor{light-gray}{gray}{0.5}

\newcommand{\eg}{{e.g.,}}

\shorttitle{The Nature of LoBAL QSOs}
\shortauthors{Lazarova et al.}

\begin{document}

\title{The Nature of LoBAL QSOs: I. SEDs and mid-infrared spectral properties}

\author{Mariana S. Lazarova\altaffilmark{1}, Gabriela Canalizo\altaffilmark{1}, Mark Lacy\altaffilmark{2}, Anna Sajina\altaffilmark{3}}

\altaffiltext{1}{Department of Physics and Astronomy, 
University of California, Riverside, CA 92521, USA;
email: mariana.lazarova@email.ucr.edu, gabriela.canalizo@ucr.edu}

\altaffiltext{2}{National ALMA Science Center, National Radio Astronomy Observatory, 520 Edgemont Rd., Charlottesville, VA 22903, USA; email: mlacy@nrao.edu}

\altaffiltext{3}{Department of Physics and Astronomy, 
Tufts University, Medford, MA 02155, USA; 
email: Anna.Sajina@tufts.edu}

\begin{abstract}

We have obtained $Spitzer$ IRS spectra and MIPS 24, 70, and 160 $\mu$m photometry for a volume-limited sample of 22 SDSS-selected Low-ionization Broad Absorption Line QSOs (LoBALs) at $0.5 < z < 0.6$. By comparing their mid-IR spectral properties and far-IR SEDs with those of a control sample of 35 non-LoBALs matched in $M_i$, we investigate the differences between the two populations in terms of their infrared emission and star formation activity. Twenty five percent of the LoBALs show PAH features and 45\% have weak 9.7$\mu$m silicate dust emission. We model the SEDs and decouple the AGN and starburst contributions to the far-infrared luminosity in LoBALs and in non-LoBALs. Their median total, starburst, and AGN infrared luminosities are comparable. Twenty percent (but no more than 60\%) of the LoBALs and 26\% of the non-LoBALs are ultra-luminous infrared galaxies (ULIRGs; L$_{IR}>10^{12}L_{\odot}$). We estimate star formation rates (SFRs) corrected for the AGN contribution to the FIR flux and find that LoBALs have comparable levels of star formation activity to non-LoBALs when considering the entire samples.  However, the SFRs of the IR-luminous LoBALs are 80\% higher than those of their counterparts in the control sample. The median contribution of star formation to the total far-infrared flux in LoBALs and in non-LoBALs is estimated to be 40-50\%, in agreement with previous results for PG QSOs. Overall, our results show that there is no strong evidence from the mid- and far-IR properties that LoBALs are drawn from a different parent population than non-LoBALs.

\end{abstract}


\section{Introduction}

Supermassive black holes are found to be ubiquitously present at the centers of galaxies with bulges and several relationships between the mass of the black hole and properties of the spheroid strongly suggest co-evolution of the two \citep[\eg][]{Kormendy1995, Magorrian1998, Laor1998, Gebhardt2000, Ferrarese2000, Kormendy2001, McLure2002, Tremaine2002}. The mechanisms via which the galaxy and black hole regulate each other's growth are still unknown. Various types of outflows, such as supernova winds and AGN-driven winds, have been invoked as the plausible feedback processes  responsible for quenching the star formation in the host and clearing the gas from the nuclear region, and, thus, halting the accretion onto the black hole and limiting its mass  \citep[\eg][]{DiMatteo2005}. However, observational evidence of the extent of their influence is still sparse and uncertain (e.g., \citealt{Moe2009}; \citealt{Bautista2010}; \citealt{Dunn2010}). 

Observations of young, recently fueled QSOs are the key to testing this evolutionary model. Particular attention has been paid to studying the properties of ultraluminous infrared galaxies (ULIRGs; $L_{IR}>10^{12}L_\odot$) since they are believed to be powered by both AGN and starbursts, although starbursts are generally believed to be responsible for the bulk of the power output (\citealt{Sanders1988}; for review see \citealt{Sanders1996}). The connection between AGN and ULIRGs is suspected by the fact that they are some of the most luminous sources in the universe with comparable luminosities of $L_{bol}> 10^{12}L_\odot$.  In addition, they are both associated with strong infrared emission from dust \citep[\eg][]{Haas2003}.  The morphologies and dynamics of ULIRGs indicate that these galaxies are associated with galaxy mergers \citep{Armus1987, Sanders1988, Murphy1996, Veilleux2002, Dasyra2006}. Similarly, many QSO hosts at $z<0.4$ show signs of interaction, even some of those that had previously been classified as ellipticals \citep{Canalizo2001, Canalizo2007, Bennert2008}. If starbursts, ULIRGs, and AGN are connected in an evolutionary sequence which was initiated by a galaxy interaction, observations of the transition stages of this process are necessary to better understand this connection \citep[see \eg][]{Hopkins2007, Hopkins2008a, Hopkins2008b}.

BAL QSOs are promising candidates for newly emerging optical QSOs. BAL QSOs are a subclass of QSOs characterized by broad absorption troughs of UV resonance lines, blueshifted relative to the QSO's rest frame, which are indicative of gas outflows with speeds of up to 0.2$c$ \citep{Foltz1983}.  BALs were rigorously defined by \citet{Weymann1991} to include only objects with broad absorption lines wider than 2000 km s$^{-1}$, blueshifted past the first 3000 km s$^{-1}$; however, some studies more recently have been more inclusive of the wide range of absorption observed and have relaxed that criterion to lower limits on the absorption width of 1000 km s$^{-1}$ \citep[\eg][]{Trump2006}. Hydrodynamic models show that AGNs are capable of launching such high velocity winds  \citep{Murray1995, Proga2000, Gallagher2012}.  Based on the material producing BAL troughs, there are at least three subclasses of BAL QSOs.  The high-ionization BAL QSOs (HiBALs) are identified via the broad absorption from \ion{C}{4} $\lambda$1549, but they might have absorption from other high-ionization species such as Ly$\alpha$, \ion{N}{5} $\lambda$1240, and \ion{Si}{4} $\lambda$1394 \citep{Hall2002}. The low-ionization BAL QSOs (LoBALs), in addition to the lines present in HiBALs, feature absorption lines from \ion{Mg}{2} $\lambda$$\lambda$ 2796,2803, \ion{Al}{3}, and \ion{Al}{2}.  A very small fraction of LoBALs, called FeLoBALs, show absorption in the rest-frame UV from metastable, excited states of \ion{Fe}{2} \citep{Hazard1987}.  

It is not well understood why only 10\%$-$30\% of the optically selected QSOs have BALs \citep{Tolea2002, Hewett2003, Trump2006, Gibson2009}, and only about one tenth of these are LoBALs \citep{Reichard2003a}.  Due to the highly obscured nature and much redder continua of these objects, optical identification omits a large fraction of BALs. Therefore, although LoBALs are observed in only 1-3\% of all optically-selected QSOs, they comprise a much higher fraction of the QSOs selected at longer wavelengths \citep{Urrutia2009, Dai2010}.  \citet{Allen2011} estimate that the intrinsic fraction of BAL QSOs can be as high as $\sim$40\% when the spectroscopic incompleteness and bias against selecting BAL QSOs in the SDSS are taken into account. Hence, BAL QSOs may only be rare in optically-selected samples.

Models attempting to explain their occurrence need to account for their obscured nature. Currently there are two competing interpretations of the BAL phenomena: orientation and evolution. On one hand, BAL and non-BAL QSOs are thought to derive from the same parent population because of the remarkable similarity in their SEDs \citep{Weymann1991, Gallagher2007}.  Nonetheless, QSO continua appear to be increasingly reddened in a sequence going from  non-BALs to HiBALs to LoBALs \citep{Reichard2003b, Richards2003}. This finding inspired efforts to explain the low occurrence of BAL QSOs within the framework of the AGN unification model (Antonucci 1993), suggesting that, due to orientation effects, BALs are seen in classical QSOs only when viewed along a narrow range of lines of sight passing through the accretion disk wind. In this picture, high column density accretion disk winds  of ionized gas are driven via resonance line absorption \citep{Murray1995, Murray1998, Elvis2000}. This model explains the low occurrence of BALs as a natural consequence of the fact that BALs are observed only at a small range of viewing angles. Although BALs are predominantly radio quiet sources \citep{Stocke1984, Stocke1992}, radio observations of the few radio-detected BALs provide a test to the orientation of the BAL wind with respect to the radio jet. Radio-detected BALs are observed at a wide range of inclinations \citep{Becker2000, Gregg2000, Brotherton2006, Montenegro-Montes2008, DiPompeo2010} suggesting that the occurrence of BALs is not a simple orientation effect \citep[e.g.,][]{DiPompeo2012}. Currently it is not clear whether or not radio-loud (RL) and radio-quiet (RQ) QSOs arise from the same parent population, so it is certainly possible that RL and RQ BAL QSOs are different classes.

An alternative model proposes that BAL QSOs are young QSOs caught during a short-lived phase in their evolution when powerful QSO-driven winds are blowing away a dusty obscuring cocoon \citep[\eg][and references therein]{Hazard1984, Voit1993, Hall2002}. This model appears to be particularly applicable to LoBALs since these objects are suspected to be young or recently refueled QSOs \citep{Boroson1992, Lipari1994} and might be exclusively associated with mergers \citep{Canalizo2001}. Observations by \citet{Canalizo2002} of the only four known LoBALs at $z < 0.4$ at the time showed that: (1) they are ULIRGs; (2) they have a small range of far-IR colors, intermediate between those characteristic of ULIRGs and QSOs; (3) their host galaxies show signs of strong tidal interactions, resulting from major mergers; (4) spectra of their hosts show unambiguous interaction-induced star formation with post-starburst ages $\leq$ 100 Myr. Similarly, studies of FeLoBALs, both at low \citep{Farrah2005} and high redshifts \citep{Farrah2007, Farrah2010}, suggest that they are associated with extremely star-forming ULIRGs.

Most recent hydrodynamic simulations of major galaxy mergers by \citet{DeBuhr2011} show that an AGN-driven BAL wind with an initial velocity $\sim$ 10000 km s$^{-1}$ would lead to a galaxy-scale outflow with velocity $\sim$ 1000 km s$^{-1}$, capable of unbinding 10$-$40\% of the initial gas of the two merging galaxies. Such AGN feedback could possibly explain the observed high-velocity outflows in post-starburst galaxies \citep{Tremonti2007} and ULIRGs \citep[\eg][]{Chung2011, Rupke2011, Sturm2011}. Further, if the paradigm suggesting that AGN feedback is responsible for regulating the growth of galaxies is correct, LoBALs may be at a unique stage where strong outflows are present, yet, star formation is still in the process of being quenched.  

Previous studies of large samples of BAL QSOs addressing their SEDs \citep{Gallagher2007} and submillimeter properties \citep{Willott2003} find that BAL and non-BAL QSOs are indistinguishable, consistent with the model that all QSOs contain BAL winds, and their detection depends on viewing angle. However, those samples mainly comprise HiBALs and refrain from drawing conclusions about LoBALs. Even if the detection of BAL troughs in QSO spectra depends on viewings angle, compelling evidence suggests that LoBALs are linked to IR-luminous galaxies, with dominant young stellar populations and disturbed morphologies. To test this possibility, we have undertaken the first multiwavelength investigation of a volume-limited sample of LoBALs. In a series of three papers, we address the nature of low-redshift LoBALs and their relationship to the broader QSO population. In particular, we test the idea that LoBALs might be a short, evolutionary stage when the AGN has been recently fueled by a merger and the ensuing winds are in the process of quenching the star formation.

In this first paper of the series, we present $Spitzer$ IRS spectroscopy and MIPS photometry at 24, 70, and 160 $\mu$m of a volume-limited, statistically-significant sample of low-redshift, optically-selected LoBALs. To study their star-forming histories, we model the infrared SEDs of LoBALs and measure their far-infrared luminosities and star formation rates. In the upcoming papers, we will study the detailed morphologies of LoBAL host galaxies via $HST$ imaging and the nature of their stellar populations via Keck LRIS spectroscopy.

Our sample of LoBALs and a control sample are described in $\S$ 2. Details of the observations are explained in $\S$ 3. We present the analysis and results in $\S$ 4. Discussion and summary of results are given in $\S$ 5. The conclusion is presented in $\S$ 6. We assume a flat universe cosmology with $H_0$ = 71 km s$^{-1}$ Mpc$^{-1}$, $\Omega_M$ = 0.27, and $\Omega_{\Lambda}$ = 0.73. All luminosities in units of the bolometric solar luminosity were calculated using $L_{\odot}$ = 3.839 $\times$ 10$^{33}$ erg s$^{-1}$.

\section{Sample selection}

\subsection{LoBALs}

We selected a volume-limited sample of 22 LoBALs from the Sloan Digital Sky Survey \citep[SDSS:][]{York2000} Third Data Release \citep[DR3;][]{Abazajian2005}. One of the spectral lines which classifies  QSOs as LoBALs is the blue-shifted broad absorption of \ion{Mg}{2} $\lambda$2800. Samples of optically-identified LoBALs are only available for $z > 0.48$, when this low-ionization broad absorption line is redshifted from rest-frame UV into the spectral range of the SDSS. While now there are available SDSS catalogues with LoBALs at $z > 0.48$ \citep[\eg][]{Gibson2009}, at the time the sample for this project was selected the most up-to-date catalog of BAL quasars was the one by \citet{Trump2006}, which identified LoBALs for $z > 0.5$. 

In addition to characterizing the FIR emission of LoBALs, we wanted to study their mid-infrared spectral properties, in particular, the PAH emission lines at rest-frame 6.2, 7.7, 8.6, 11.3, and 12.8 $\mu$m and the silicate emission at rest-frame 9.7 $\mu$m. Choosing objects at redshifts less than 0.6 ensures that these spectral lines are observable within the wavelength range 7$ - $21 $\mu$m, allowing us to  use only two of the four channels of the $Spitzer$ infrared spectrograph (i.e., SL1 + LL2; see Section \ref{observations} for details). An upper limit of $z < $ 0.6 also makes it feasible to resolve the host galaxy morphologies on $HST$ images despite the bright nuclear emission. An attempt to study the morphologies of LoBALs at $0.9 < z < 2.0$ showed that the host galaxies could not be resolved at those redshifts (i.e., Fan; $HST$ PID 10237), which strongly motivated us to choose a lower redshift upper limit.

With these considerations in mind, our sample was selected to include all LoBALs within the redshift range 0.5 $< z <$ 0.6 from the \citet{Trump2006} catalog of BAL quasars,  drawn from Data Release 3 of the Sloan Digital Sky Survey quasar catalog by \citet{Schneider2005}. In the \citet{Trump2006} catalog, QSOs with regions of flux at least 10\% below the continuum, spanning over a velocity range of at least 1000 km s$^{-1}$ blue-ward of Mg~{\sc II} $\lambda$2800, were identified as LoBALs. They found 457 LoBALs in the redshift range 0.5 $< z <$ 2.15. Of those, only 22 fall within 0.5 $< z <$ 0.6, when we exclude one object which is classified as a narrow-line LoBAL and one identified as an uncertain FeLoBAL. The volume-limited sample of 22 low-redshift LoBALs is listed in Table~\ref{table:sdss}.  Note that some of the objects in our sample are not identified as LoBALs in the catalog by \citet{Gibson2009} since the Gibson et al.\ introduce a new balnicity index, which is different from the absorption index of \citet{Trump2006}.   The SDSS spectra of the objects in our sample, showing the \ion{Mg}{2} absorption trough, are included in the Appendix.

\subsection{Control sample of non-LoBAL type-1 QSOs}

In order to investigate the relationship between LoBALs and classical QSOs, we compiled a sample of type-1 QSOs with sufficient available archival data to be analyzed in the same way as the LoBAL sample. We selected objects whose spectra did not show Mg~{\sc II}  BAL absorption. Since we do not have UV spectra to determine whether any of these objects are HiBALs, we shall refer to them as "non-LoBALs." We drew the control sample from the compilation of quasar SEDs by \citet{Richards2006}, who published all available SDSS, $Spitzer$ MIPS, and IRAC photometry, as well as x-ray and radio data, of the type-1 quasars in the SDSS-DR3 quasar catalog by \citet{Schneider2005}.  It is important to match the LoBALs to type-1 QSOs of comparable luminosity.  Limiting the control sample to the same redshift range as the LoBALs, i.e., 0.5 $< z <$ 0.6, selected only 11 non-LoBALs whose absolute $M_i$ magnitudes matched only the lower luminosity LoBALs.  To avoid the problem of having an under-luminous control sample, we matched the control sample to the reddening-corrected optical luminosity of the LoBALs, not restricting the redshift.  Moreover, only few LoBALs have been found to have radio emission \citep[\eg][]{Becker1997, Becker2000, Becker2001, Brotherton1998, Menou2001}, which limits our selection to the radio-quiet sub-population.

In summary, we selected all radio-quiet non-LoBALs  from \citet{Richards2006}, which matched the absolute $M_i$ magnitude range of our LoBAL sample,  $-22.4 > M_i >-$25.6 (see Table~\ref{table:sdss}).  A total of 35 objects within 0.45 $< z <$ 0.83 satisfied these criteria: 16 from the Lockman hole, six from the ELAIS-N1, six from the ELAIS-N2 fields in the $Spitzer$ Wide-Area Infrared Extragalactic Survey \citep[SWIRE:][]{Lonsdale2003}, and seven from the $Spitzer$ observations of the Extragalactic First Look Survey area \citep[xFLS:][]{Frayer2006}. Tables~\ref{table:sdss},~\ref{table:ir} and~\ref{table:controlirac} list relevant optical and infrared photometry of the comparison sample of non-LoBALs.

\begin{table*}
\scriptsize
\caption{SDSS DR7 PSF magnitudes corrected for the given here Galactic extinction.}
\centering
\begin{tabular}{cccccccccccc}
\hline\hline
\#	&	SDSS Object ID	&	Redshift	&	$M_i$	&	E($B-V$)	&	u			&	g				&	r			&	i			&	z			\\
	&		&		&	[AB mag]	&	[AB mag]	&	[AB mag]			&	[AB mag]				&	[AB mag]			&	[AB mag]			&	[AB mag]			\\
\hline\hline	
 \multicolumn{10}{c}{\bf\footnotesize LoBAL sample}  \\
\hline\hline																													
1	&	J023102.49$-$083141.2	&	0.596	&	-23.54	&	0.04	&	19.54	$\pm$	0.04	&	19.17	$\pm$	0.01		&	19.16	$\pm$	0.02	&	18.96	$\pm$	0.02	&	18.93	$\pm$	0.06	\\
2	&	J023153.63$-$093333.5	&	0.587	&	-23.39	&	0.03	&	19.95	$\pm$	0.05	&	19.37	$\pm$	0.01		&	19.24	$\pm$	0.01	&	18.96	$\pm$	0.02	&	18.96	$\pm$	0.05	\\
3	&	J025026.66+000903.4	&	0.554	&	-24.05	&	0.07	&	20.90	$\pm$	0.13	&	19.84	$\pm$	0.02		&	19.10	$\pm$	0.01	&	18.57	$\pm$	0.01	&	18.46	$\pm$	0.04	\\
4	&	J083525.98+435211.2	&	0.568	&	-25.14	&	0.03	&	18.31	$\pm$	0.02	&	17.48	$\pm$	0.01		&	17.47	$\pm$	0.01	&	17.27	$\pm$	0.01	&	17.32	$\pm$	0.01	\\
5	&	J085053.12+445122.5	&	0.541	&	-25.11	&	0.03	&	17.84	$\pm$	0.01	&	17.44	$\pm$	0.01		&	17.33	$\pm$	0.01	&	17.19	$\pm$	0.01	&	17.27	$\pm$	0.01	\\
6	&	J085215.66+492040.8	&	0.566	&	-23.82	&	0.02	&	19.37	$\pm$	0.03	&	18.94	$\pm$	0.01		&	18.77	$\pm$	0.01	&	18.57	$\pm$	0.01	&	18.57	$\pm$	0.03	\\
7	&	J085357.87+463350.6	&	0.550	&	-24.43	&	0.02	&	18.67	$\pm$	0.02	&	18.18	$\pm$	0.01		&	18.19	$\pm$	0.01	&	17.89	$\pm$	0.01	&	17.83	$\pm$	0.02	\\
8	&	J101151.95+542942.7	&	0.536	&	-23.32	&	0.01	&	20.08	$\pm$	0.04	&	19.60	$\pm$	0.02		&	19.34	$\pm$	0.01	&	18.91	$\pm$	0.01	&	19.04	$\pm$	0.04	\\
9	&	J102802.32+592906.6	&	0.535	&	-23.47	&	0.01	&	18.94	$\pm$	0.02	&	18.76	$\pm$	0.01		&	18.89	$\pm$	0.01	&	18.75	$\pm$	0.01	&	18.82	$\pm$	0.06	\\
10	&	J105102.77+525049.8	&	0.543	&	-23.40	&	0.01	&	19.88	$\pm$	0.05	&	19.31	$\pm$	0.01		&	19.24	$\pm$	0.01	&	18.87	$\pm$	0.01	&	18.85	$\pm$	0.04	\\
11	&	J105404.73+042939.3	&	0.578	&	-23.47	&	0.04	&	20.20	$\pm$	0.05	&	19.55	$\pm$	0.01		&	19.43	$\pm$	0.01	&	19.00	$\pm$	0.01	&	19.05	$\pm$	0.04	\\
12	&	J112822.41+482309.9	&	0.543	&	-25.05	&	0.02	&	18.51	$\pm$	0.02	&	17.83	$\pm$	0.01		&	17.56	$\pm$	0.01	&	17.23	$\pm$	0.01	&	17.17	$\pm$	0.01	\\
13	&	J114043.62+532439.0	&	0.530	&	-24.07	&	0.01	&	18.73	$\pm$	0.02	&	18.32	$\pm$	0.01		&	18.32	$\pm$	0.01	&	18.14	$\pm$	0.01	&	18.13	$\pm$	0.02	\\
14	&	J130952.89+011950.6	&	0.547	&	-25.12	&	0.04	&	17.79	$\pm$	0.01	&	17.47	$\pm$	0.01		&	17.41	$\pm$	0.01	&	17.21	$\pm$	0.01	&	17.20	$\pm$	0.01	\\
15	&	J140025.53$-$012957.0	&	0.584	&	-24.43	&	0.05	&	19.49	$\pm$	0.04	&	18.49	$\pm$	0.01		&	18.30	$\pm$	0.01	&	18.09	$\pm$	0.01	&	18.10	$\pm$	0.03	\\
16	&	J141946.36+463424.3	&	0.546	&	-22.41	&	0.01	&	20.89	$\pm$	0.07	&	20.44	$\pm$	0.02		&	20.33	$\pm$	0.02	&	19.87	$\pm$	0.02	&	19.48	$\pm$	0.05	\\
17	&	J142649.24+032517.7	&	0.530	&	-24.18	&	0.04	&	18.80	$\pm$	0.03	&	18.37	$\pm$	0.01		&	18.33	$\pm$	0.01	&	18.08	$\pm$	0.01	&	18.01	$\pm$	0.02	\\
18	&	J142927.28+523849.5	&	0.594	&	-25.29	&	0.01	&	18.03	$\pm$	0.01	&	17.56	$\pm$	0.01		&	17.48	$\pm$	0.01	&	17.19	$\pm$	0.01	&	17.30	$\pm$	0.01	\\
19	&	J161425.17+375210.7	&	0.553	&	-25.55	&	0.02	&	17.24	$\pm$	0.01	&	16.92	$\pm$	0.00		&	16.92	$\pm$	0.00	&	16.77	$\pm$	0.01	&	16.89	$\pm$	0.01	\\
20	&	J170010.83+395545.8	&	0.577	&	-23.40	&	0.02	&	19.80	$\pm$	0.04	&	19.45	$\pm$	0.01		&	19.26	$\pm$	0.01	&	19.03	$\pm$	0.01	&	18.91	$\pm$	0.05	\\
21	&	J170341.82+383944.7	&	0.554	&	-24.14	&	0.04	&	19.95	$\pm$	0.04	&	19.08	$\pm$	0.01		&	18.65	$\pm$	0.01	&	18.23	$\pm$	0.01	&	18.17	$\pm$	0.03	\\
22	&	J204333.20$-$001104.2	&	0.545	&	-24.82	&	0.06	&	18.43	$\pm$	0.02	&	18.01	$\pm$	0.01		&	17.79	$\pm$	0.01	&	17.56	$\pm$	0.01	&	17.64	$\pm$	0.02	\\
\hline\hline
 \multicolumn{10}{c}{\bf \footnotesize Control sample of non-LoBALs}  \\
\hline\hline						
1	&	J103236.22+580033.9	&	0.687	&	-23.20	&	0.006	&	20.23	$\pm$	0.05	&	19.85	$\pm$	0.03	&	19.82	$\pm$	0.03	&	19.63	$\pm$	0.03	&	19.79	$\pm$	0.08	\\
2	&	J103333.92+582818.8	&	0.574	&	-22.17	&	0.007	&	20.78	$\pm$	0.07	&	20.29	$\pm$	0.03	&	20.40	$\pm$	0.04	&	20.22	$\pm$	0.05	&	20.06	$\pm$	0.14	\\
3	&	J103651.94+575950.9	&	0.500	&	-23.45	&	0.006	&	19.48	$\pm$	0.03	&	19.05	$\pm$	0.02	&	18.79	$\pm$	0.02	&	18.60	$\pm$	0.02	&	18.50	$\pm$	0.04	\\
4	&	J103721.15+590755.7	&	0.603	&	-23.03	&	0.008	&	20.04	$\pm$	0.15	&	19.43	$\pm$	0.03	&	19.59	$\pm$	0.02	&	19.49	$\pm$	0.02	&	19.67	$\pm$	0.10	\\
5	&	J104210.25+594253.5	&	0.675	&	-23.17	&	0.012	&	19.85	$\pm$	0.09	&	19.66	$\pm$	0.04	&	19.65	$\pm$	0.02	&	19.63	$\pm$	0.04	&	19.70	$\pm$	0.12	\\
6	&	J104526.73+595422.6	&	0.646	&	-23.85	&	0.011	&	19.44	$\pm$	0.05	&	19.06	$\pm$	0.03	&	19.14	$\pm$	0.02	&	18.83	$\pm$	0.03	&	18.84	$\pm$	0.05	\\
7	&	J104556.84+570747.0	&	0.541	&	-23.24	&	0.007	&	19.21	$\pm$	0.03	&	18.95	$\pm$	0.02	&	19.06	$\pm$	0.02	&	19.00	$\pm$	0.03	&	18.99	$\pm$	0.04	\\
8	&	J104625.02+584839.1	&	0.577	&	-23.70	&	0.010	&	19.06	$\pm$	0.02	&	18.76	$\pm$	0.03	&	18.84	$\pm$	0.03	&	18.70	$\pm$	0.02	&	18.78	$\pm$	0.04	\\
9	&	J104633.70+571530.4	&	0.712	&	-24.16	&	0.008	&	19.03	$\pm$	0.02	&	18.72	$\pm$	0.01	&	18.77	$\pm$	0.02	&	18.75	$\pm$	0.01	&	18.76	$\pm$	0.03	\\
10	&	J104840.28+563635.6	&	0.714	&	-23.86	&	0.007	&	19.67	$\pm$	0.04	&	19.25	$\pm$	0.03	&	19.20	$\pm$	0.02	&	19.06	$\pm$	0.02	&	18.90	$\pm$	0.05	\\
11	&	J104857.92+560112.3	&	0.800	&	-24.25	&	0.009	&	19.15	$\pm$	0.02	&	18.81	$\pm$	0.04	&	18.82	$\pm$	0.01	&	18.95	$\pm$	0.02	&	18.77	$\pm$	0.04	\\
12	&	J105000.21+581904.2	&	0.833	&	-25.53	&	0.008	&	17.87	$\pm$	0.01	&	17.67	$\pm$	0.02	&	17.64	$\pm$	0.02	&	17.77	$\pm$	0.01	&	17.68	$\pm$	0.03	\\
13	&	J105106.12+591625.1	&	0.768	&	-24.78	&	0.009	&	18.60	$\pm$	0.03	&	18.22	$\pm$	0.03	&	18.27	$\pm$	0.02	&	18.33	$\pm$	0.02	&	18.19	$\pm$	0.03	\\
14	&	J105518.08+570423.5	&	0.696	&	-24.13	&	0.007	&	18.87	$\pm$	0.02	&	18.59	$\pm$	0.02	&	18.71	$\pm$	0.01	&	18.73	$\pm$	0.02	&	18.68	$\pm$	0.03	\\
15	&	J105604.00+581523.4	&	0.832	&	-24.36	&	0.007	&	19.02	$\pm$	0.03	&	18.89	$\pm$	0.02	&	18.81	$\pm$	0.03	&	18.93	$\pm$	0.02	&	18.81	$\pm$	0.03	\\
16	&	J105959.93+574848.1	&	0.453	&	-23.76	&	0.006	&	18.86	$\pm$	0.03	&	18.36	$\pm$	0.03	&	18.25	$\pm$	0.02	&	18.05	$\pm$	0.02	&	17.81	$\pm$	0.02	\\
17	&	J160015.68+552259.9	&	0.673	&	-24.36	&	0.007	&	18.90	$\pm$	0.03	&	18.54	$\pm$	0.02	&	18.54	$\pm$	0.01	&	18.42	$\pm$	0.02	&	18.36	$\pm$	0.04	\\
18	&	J160128.54+544521.3	&	0.728	&	-24.93	&	0.010	&	18.37	$\pm$	0.02	&	18.12	$\pm$	0.02	&	18.06	$\pm$	0.02	&	18.05	$\pm$	0.02	&	18.00	$\pm$	0.03	\\
19	&	J160341.44+541501.5	&	0.580	&	-23.23	&	0.008	&	19.61	$\pm$	0.05	&	19.26	$\pm$	0.02	&	19.30	$\pm$	0.03	&	19.19	$\pm$	0.03	&	19.14	$\pm$	0.05	\\
20	&	J160523.10+545613.3	&	0.572	&	-23.63	&	0.009	&	19.12	$\pm$	0.04	&	18.84	$\pm$	0.04	&	18.91	$\pm$	0.03	&	18.76	$\pm$	0.02	&	18.81	$\pm$	0.05	\\
21	&	J160630.60+542007.5	&	0.820	&	-24.43	&	0.008	&	18.97	$\pm$	0.03	&	18.75	$\pm$	0.03	&	18.72	$\pm$	0.02	&	18.83	$\pm$	0.03	&	18.61	$\pm$	0.04	\\
22	&	J160908.95+533153.2	&	0.816	&	-24.57	&	0.010	&	19.01	$\pm$	0.03	&	18.48	$\pm$	0.03	&	18.48	$\pm$	0.04	&	18.68	$\pm$	0.02	&	18.50	$\pm$	0.03	\\
23	&	J163031.46+410145.6	&	0.531	&	-23.48	&	0.008	&	18.90	$\pm$	0.02	&	18.69	$\pm$	0.02	&	18.87	$\pm$	0.01	&	18.72	$\pm$	0.02	&	18.66	$\pm$	0.03	\\
24	&	J163135.46+405756.4	&	0.750	&	-24.19	&	0.009	&	19.48	$\pm$	0.03	&	19.00	$\pm$	0.01	&	18.85	$\pm$	0.02	&	18.86	$\pm$	0.02	&	18.67	$\pm$	0.04	\\
25	&	J163143.76+404735.6	&	0.538	&	-23.55	&	0.008	&	19.57	$\pm$	0.03	&	19.11	$\pm$	0.02	&	18.95	$\pm$	0.02	&	18.68	$\pm$	0.02	&	18.65	$\pm$	0.03	\\
26	&	J163352.34+402115.5	&	0.782	&	-24.05	&	0.007	&	19.21	$\pm$	0.03	&	18.86	$\pm$	0.02	&	18.95	$\pm$	0.02	&	19.10	$\pm$	0.02	&	18.88	$\pm$	0.05	\\
27	&	J163502.80+412952.9	&	0.472	&	-23.97	&	0.006	&	18.15	$\pm$	0.02	&	17.94	$\pm$	0.02	&	17.97	$\pm$	0.01	&	17.94	$\pm$	0.02	&	17.91	$\pm$	0.02	\\
28	&	J163854.62+415419.5	&	0.711	&	-24.30	&	0.009	&	19.05	$\pm$	0.05	&	18.78	$\pm$	0.02	&	18.68	$\pm$	0.02	&	18.62	$\pm$	0.02	&	18.60	$\pm$	0.03	\\
29	&	J171126.94+585544.2	&	0.537	&	-23.49	&	0.024	&	19.16	$\pm$	0.03	&	18.80	$\pm$	0.02	&	18.90	$\pm$	0.02	&	18.77	$\pm$	0.02	&	18.72	$\pm$	0.05	\\
30	&	J171334.02+595028.3	&	0.615	&	-24.62	&	0.021	&	18.16	$\pm$	0.02	&	17.88	$\pm$	0.01	&	18.00	$\pm$	0.02	&	17.96	$\pm$	0.02	&	18.13	$\pm$	0.03	\\
31	&	J171736.90+593011.4	&	0.600	&	-23.78	&	0.021	&	19.21	$\pm$	0.03	&	18.81	$\pm$	0.02	&	18.90	$\pm$	0.02	&	18.75	$\pm$	0.02	&	18.83	$\pm$	0.05	\\
32	&	J171748.43+594820.6	&	0.763	&	-25.02	&	0.022	&	18.27	$\pm$	0.03	&	17.99	$\pm$	0.02	&	18.02	$\pm$	0.02	&	18.10	$\pm$	0.02	&	18.05	$\pm$	0.03	\\
33	&	J171818.14+584905.2	&	0.634	&	-24.18	&	0.028	&	19.00	$\pm$	0.03	&	18.59	$\pm$	0.02	&	18.60	$\pm$	0.02	&	18.50	$\pm$	0.02	&	18.64	$\pm$	0.04	\\
34	&	J172104.75+592451.4	&	0.786	&	-23.85	&	0.028	&	19.62	$\pm$	0.04	&	19.39	$\pm$	0.02	&	19.32	$\pm$	0.02	&	19.35	$\pm$	0.03	&	19.13	$\pm$	0.06	\\
35	&	J172414.05+593644.0	&	0.745	&	-24.29	&	0.025	&	19.33	$\pm$	0.03	&	18.90	$\pm$	0.02	&	18.80	$\pm$	0.01	&	18.77	$\pm$	0.02	&	18.55	$\pm$	0.03	\\	
\hline                           
\end{tabular}      
\label{table:sdss}             
\end{table*}

\begin{table*}
\scriptsize
\caption{2MASS and $Spitzer$ MIPS infrared photometry.}
\centering
\begin{tabular}{cccccccc}
\hline\hline

\#	&	SDSS Object ID	&	$J$			&	$H$			&	$K_s$			&	$f_{24}$			&		$f_{70}$			&		$f_{160}$			\\
	&		&	[Vega mag]			&	[Vega mag]			&	[Vega mag]			&	[mJy]			&		[mJy]			&		[mJy]			\\
\hline\hline
\multicolumn{8}{c}{\bf\footnotesize LoBAL sample}\\	
\hline\hline				
1	&	J023102.49$-$083141.2	&		$\cdots$		&		$\cdots$		&	15.71	$\pm$	0.23	&	5.04	$\pm$	0.20	&	$<$	19.63			&	$<$	29.30			\\
2	&	J023153.63$-$093333.5	&		$\cdots$		&		$\cdots$		&	15.92	$\pm$	0.25	&	1.17	$\pm$	0.17	&	$<$	20.55			&	$<$	22.21			\\
3	&	J025026.66+000903.4	&	17.10	$\pm$	0.18	&		$\cdots$		&	15.42	$\pm$	0.15	&	10.18	$\pm$	0.20	&		86.66	$\pm$	6.68	&		86.64	$\pm$	15.02	\\
4	&	J083525.98+435211.2	&	16.43	$\pm$	0.13	&	15.98	$\pm$	0.20	&	15.14	$\pm$	0.12	&	$\cdots$			&		$\cdots$			&		$\cdots$			\\
5	&	J085053.12+445122.5	&	16.24	$\pm$	0.11	&	15.98	$\pm$	0.20	&	14.92	$\pm$	0.12	&	5.68	$\pm$	0.16	&	$<$	17.56			&	$<$	23.17			\\
6	&	J085215.66+492040.8	&		$\cdots$		&		$\cdots$		&		$\cdots$		&	1.30	$\pm$	0.16	&	$<$	17.07			&	$<$	21.88			\\
7	&	J085357.87+463350.6	&	16.78	$\pm$	0.17	&		$\cdots$		&	15.29	$\pm$	0.15	&	1.90	$\pm$	0.16	&	$<$	16.08			&	$<$	20.78			\\
8	&	J101151.95+542942.7	&		$\cdots$		&		$\cdots$		&		$\cdots$		&	4.52	$\pm$	0.16	&		49.81	$\pm$	7.18	&		37.44	$\pm$	7.98	\\
9	&	J102802.32+592906.6	&		$\cdots$		&		$\cdots$		&		$\cdots$		&	$\cdots$			&		$\cdots$			&		$\cdots$			\\
10	&	J105102.77+525049.8	&	17.31	$\pm$	0.23	&		$\cdots$		&	15.19	$\pm$	0.12	&	3.28	$\pm$	0.15	&	$<$	18.41			&	$<$	17.80			\\
11	&	J105404.73+042939.3	&		$\cdots$		&		$\cdots$		&		$\cdots$		&	$\cdots$			&		$\cdots$			&		$\cdots$			\\
12	&	J112822.41+482309.9	&	16.07	$\pm$	0.12	&	15.39	$\pm$	0.14	&	14.45	$\pm$	0.09	&	$\cdots$			&		$\cdots$			&		$\cdots$			\\
13	&	J114043.62+532439.0	&	16.95	$\pm$	0.20	&		$\cdots$		&	15.31	$\pm$	0.17	&	$\cdots$			&		$\cdots$			&		$\cdots$			\\
14	&	J130952.89+011950.6	&	15.93	$\pm$	0.09	&	15.34	$\pm$	0.11	&	14.61	$\pm$	0.10	&	10.21	$\pm$	0.24	&	$<$	20.93			&	$<$	27.58			\\
15	&	J140025.53$-$012957.0	&		$\cdots$		&		$\cdots$		&		$\cdots$		&	1.79	$\pm$	0.20	&	$<$	28.43			&	$<$	30.49			\\
16	&	J141946.36+463424.3	&		$\cdots$		&		$\cdots$		&		$\cdots$		&	2.02	$\pm$	0.14	&	$<$	17.10			&		54.75	$\pm$	7.75	\\
17	&	J142649.24+032517.7	&	16.75	$\pm$	0.19	&	16.03	$\pm$	0.17	&	15.19	$\pm$	0.17	&	5.85	$\pm$	0.20	&	$<$	20.50			&		$\cdots$			\\
18	&	J142927.28+523849.5	&	16.27	$\pm$	0.10	&	15.59	$\pm$	0.12	&	14.85	$\pm$	0.09	&	8.84	$\pm$	0.17	&	$<$	14.17			&	$<$	25.28			\\
19	&	J161425.17+375210.7	&	16.10	$\pm$	0.08	&	15.44	$\pm$	0.12	&	14.38	$\pm$	0.08	&	20.06	$\pm$	0.21	&		110.26	$\pm$	6.63	&		$\cdots$			\\
20	&	J170010.83+395545.8	&		$\cdots$		&		$\cdots$		&	15.65	$\pm$	0.21	&	4.91	$\pm$	0.16	&		45.32	$\pm$	6.43	&		$\cdots$			\\
21	&	J170341.82+383944.7	&	16.82	$\pm$	0.17	&	16.02	$\pm$	0.20	&	15.44	$\pm$	0.17	&	8.83	$\pm$	0.17	&	$<$	17.54			&		$\cdots$			\\
22	&	J204333.20$-$001104.2	&	16.84	$\pm$	0.15	&	15.80	$\pm$	0.15	&	15.24	$\pm$	0.15	&	5.31	$\pm$	0.18	&	$<$	15.23			&		$\cdots$			\\
\hline\hline
\multicolumn{8}{c}{\bf\footnotesize Control sample of non-LoBALs}\\	
\hline\hline
1	&	103236.22+580033.9	&	18.46		0.26	&	17.59		0.28	&		$\cdots$	0.00	&	1.16	$\pm$	0.02	&	$<$	4.41	$\pm$	1.47	&	$<$	12.46	$\pm$	4.15	\\
2	&	103333.92+582818.8	&		$\cdots$		&		$\cdots$		&	16.57	$\pm$	0.15	&	0.74	$\pm$	0.02	&	$<$	4.52	$\pm$	1.51	&	$<$	13.12	$\pm$	4.37	\\
3	&	103651.94+575950.9	&	16.96	$\pm$	0.19	&	16.05	$\pm$	0.20	&	15.40	$\pm$	0.18	&	4.59	$\pm$	0.02	&		27.97	$\pm$	2.14	&	$<$	12.19	$\pm$	4.06	\\
4	&	103721.15+590755.7	&	18.70	$\pm$	0.23	&		$\cdots$		&	16.90	$\pm$	0.20	&	2.13	$\pm$	0.02	&	$<$	4.36	$\pm$	1.45	&	$<$	14.41	$\pm$	4.80	\\
5	&	104210.25+594253.5	&	18.45	$\pm$	0.24	&		$\cdots$		&		$\cdots$		&	1.08	$\pm$	0.02	&	$<$	3.92	$\pm$	1.31	&	$<$	17.06	$\pm$	5.69	\\
6	&	104526.73+595422.6	&	17.78	$\pm$	0.17	&	16.64	$\pm$	0.14	&	16.11	$\pm$	0.15	&	4.38	$\pm$	0.02	&		8.20	$\pm$	2.53	&	$<$	12.48	$\pm$	4.16	\\
7	&	104556.84+570747.0	&	17.86	$\pm$	0.14	&	16.94	$\pm$	0.18	&	16.28	$\pm$	0.15	&	2.72	$\pm$	0.02	&	$<$	4.46	$\pm$	1.49	&	$<$	11.76	$\pm$	3.92	\\
8	&	104625.02+584839.1	&	17.97	$\pm$	0.17	&	16.87	$\pm$	0.19	&	16.05	$\pm$	0.13	&	3.88	$\pm$	0.02	&	$<$	3.35	$\pm$	1.12	&	$<$	7.79	$\pm$	2.60	\\
9	&	104633.70+571530.4	&	17.57	$\pm$	0.12	&		$\cdots$		&	16.10	$\pm$	0.13	&	1.84	$\pm$	0.02	&	$<$	4.75	$\pm$	1.58	&	$<$	9.52	$\pm$	3.17	\\
10	&	104840.28+563635.6	&	17.80	$\pm$	0.14	&	16.85	$\pm$	0.15	&	16.42	$\pm$	0.19	&	2.66	$\pm$	0.02	&	$<$	5.78	$\pm$	1.93	&	$<$	11.93	$\pm$	3.98	\\
11	&	104857.92+560112.3	&	17.55	$\pm$	0.13	&		$\cdots$		&	16.54	$\pm$	0.23	&	1.37	$\pm$	0.02	&		27.09	$\pm$	2.05	&	$<$	12.11	$\pm$	4.04	\\
12	&	105000.21+581904.2	&	16.80	$\pm$	0.16	&	16.20	$\pm$	0.24	&	15.46	$\pm$	0.18	&	4.62	$\pm$	0.02	&	$<$	4.62	$\pm$	1.54	&	$<$	12.93	$\pm$	4.31	\\
13	&	105106.12+591625.1	&	17.17	$\pm$	0.11	&	16.62	$\pm$	0.13	&	15.34	$\pm$	0.15	&	5.39	$\pm$	0.02	&		23.33	$\pm$	2.15	&	$<$	10.41	$\pm$	3.47	\\
14	&	105518.08+570423.5	&	17.61	$\pm$	0.16	&		$\cdots$		&	16.36	$\pm$	0.17	&	3.55	$\pm$	0.02	&	$<$	5.49	$\pm$	1.83	&	$<$	9.51	$\pm$	3.17	\\
15	&	105604.00+581523.4	&	17.28	$\pm$	0.13	&	16.87	$\pm$	0.21	&	16.23	$\pm$	0.22	&	4.09	$\pm$	0.02	&	$<$	3.70	$\pm$	1.23	&	$<$	11.50	$\pm$	3.83	\\
16	&	105959.93+574848.1	&	16.65	$\pm$	0.16	&	16.08	$\pm$	0.22	&	15.48	$\pm$	0.16	&	9.04	$\pm$	0.03	&		17.31	$\pm$	2.45	&	$<$	11.23	$\pm$	3.74	\\
17	&	160015.68+552259.9	&		$\cdots$		&		$\cdots$		&	15.70	$\pm$	0.22	&	4.48	$\pm$	0.02	&		17.3733	$\pm$	2.54	&	$<$	11.66	$\pm$	3.89	\\
18	&	160128.54+544521.3	&	16.78	$\pm$	0.24	&		$\cdots$		&	16.13	$\pm$	0.30	&	12.93	$\pm$	0.02	&		38.14	$\pm$	2.57	&	$<$	11.68	$\pm$	3.89	\\
19	&	160341.44+541501.5	&	17.22	$\pm$	0.33	&		$\cdots$		&		$\cdots$		&	1.18	$\pm$	0.02	&	$<$	6.78	$\pm$	2.26	&	$<$	11.01	$\pm$	3.67	\\
20	&	160523.10+545613.3	&		$\cdots$		&		$\cdots$		&	16.07	$\pm$	0.27	&	4.74	$\pm$	0.02	&		7.88	$\pm$	1.81	&	$<$	10.78	$\pm$	3.59	\\
21	&	160630.60+542007.5	&		$\cdots$		&		$\cdots$		&		$\cdots$		&	5.56	$\pm$	0.02	&	$<$	4.77	$\pm$	1.59	&	$<$	10.17	$\pm$	3.39	\\
22	&	160908.95+533153.2	&		$\cdots$		&	16.22	$\pm$	0.20	&		$\cdots$		&	3.03	$\pm$	0.02	&	$<$	4.92	$\pm$	1.64	&	$<$	11.28	$\pm$	3.76	\\
23	&	163031.46+410145.6	&		$\cdots$		&		$\cdots$		&	15.69	$\pm$	0.24	&	2.17	$\pm$	0.02	&		12.03	$\pm$	1.69	&	$<$	4.47	$\pm$	1.49	\\
24	&	163135.46+405756.4	&	17.33	$\pm$	0.32	&	16.21	$\pm$	0.24	&		$\cdots$		&	4.3	$\pm$	0.02	&		10.98	$\pm$	1.99	&	$<$	4.57	$\pm$	1.52	\\
25	&	163143.76+404735.6	&	17.13	$\pm$	0.28	&		$\cdots$		&	15.96	$\pm$	0.27	&	3.9	$\pm$	0.02	&		13.91	$\pm$	1.88	&	$<$	4.71	$\pm$	1.57	\\
26	&	163352.34+402115.5	&	17.16	$\pm$	0.27	&		$\cdots$		&		$\cdots$		&	2.86	$\pm$	0.02	&		13.88	$\pm$	2.23	&	$<$	4.56	$\pm$	1.52	\\
27	&	163502.80+412952.9	&	17.08	$\pm$	0.22	&	16.30	$\pm$	0.21	&	15.22	$\pm$	0.16	&	3.92	$\pm$	0.02	&	$<$	5.13	$\pm$	1.71	&	$<$	5.30	$\pm$	1.77	\\
28	&	163854.62+415419.5	&	17.41	$\pm$	0.33	&		$\cdots$		&		$\cdots$		&	2.91	$\pm$	0.02	&	$<$	4.21	$\pm$	1.40	&	$<$	4.54	$\pm$	1.51	\\
29	&	171126.94+585544.2	&		$\cdots$		&		$\cdots$		&		$\cdots$		&	3.45	$\pm$	0.07	&		29.10	$\pm$	2.24	&		98.32	$\pm$	11.27	\\
30	&	171334.02+595028.3	&	17.12	$\pm$	0.27	&	16.50	$\pm$	0.24	&	15.47	$\pm$	0.20	&	5.38	$\pm$	0.07	&	$<$	1.99	$\pm$	0.66	&	$<$	20.21	$\pm$	6.74	\\
31	&	171736.90+593011.4	&	17.63	$\pm$	0.35	&		$\cdots$		&		$\cdots$		&	6.38	$\pm$	0.05	&		19.81	$\pm$	1.46	&	$<$	12.33	$\pm$	4.11	\\
32	&	171748.43+594820.6	&	17.20	$\pm$	0.33	&		$\cdots$		&		$\cdots$		&	3.04	$\pm$	0.04	&		10.34	$\pm$	1.30	&	$<$	7.72	$\pm$	2.57	\\
33	&	171818.14+584905.2	&		$\cdots$		&		$\cdots$		&		$\cdots$		&	4.06	$\pm$	0.06	&		30.43	$\pm$	2.26	&	$<$	7.05	$\pm$	2.35	\\
34	&	172104.75+592451.4	&	17.39	$\pm$	0.32	&		$\cdots$		&	16.19	$\pm$	0.29	&	4.44	$\pm$	0.06	&		12.56	$\pm$	2.10	&	$<$	18.29	$\pm$	6.10	\\
35	&	172414.05+593644.0	&		$\cdots$		&		$\cdots$		&	15.75	$\pm$	0.24	&	1.81	$\pm$	0.06	&	$<$	3.58	$\pm$	1.19	&	$<$	16.80	$\pm$	5.60	\\
\hline
\end{tabular}               
\label{table:ir}     
\end{table*}

\begin{table*}
\footnotesize
\caption{Control sample of non-LoBALs: MIPS data fields and archival IRAC photometry.}
\centering
\begin{tabular}{ccccccc}					
\hline\hline																					
\#	&	SDSS Object ID	&	MIPS field	&	$f_{3.6}$			&	$f_{4.5}$			&	$f_{5.8}$			&	$f_{8.0}$			\\
	&		&		&	($\mu$Jy)			&	($\mu$Jy)			&	($\mu$Jy)			&	($\mu$Jy)			\\
\hline\hline																					
1	&	103236.22+580033.9	&	SWIRE$-$Lockman Hole	&	168.4	$\pm$	1.0	&	245.0	$\pm$	1.3	&	342.8	$\pm$	4.7	&	492.8	$\pm$	4.7	\\
2	&	103333.92+582818.8	&	SWIRE$-$Lockman Hole	&	170.2	$\pm$	1.4	&	188.0	$\pm$	2.0	&	233.4	$\pm$	5.4	&	247.0	$\pm$	5.7	\\
3	&	103651.94+575950.9	&	SWIRE$-$Lockman Hole	&	694.2	$\pm$	3.6	&	864.5	$\pm$	4.2	&	1079.7	$\pm$	11.0	&	1368.0	$\pm$	8.3	\\
4	&	103721.15+590755.7	&	SWIRE$-$Lockman Hole	&	249.7	$\pm$	1.6	&	353.6	$\pm$	2.7	&	520.3	$\pm$	7.7	&	764.4	$\pm$	7.0	\\
5	&	104210.25+594253.5	&	SWIRE$-$Lockman Hole	&	250.3	$\pm$	2.1	&	308.6	$\pm$	1.8	&	375.1	$\pm$	7.0	&	479.9	$\pm$	4.6	\\
6	&	104526.73+595422.6	&	SWIRE$-$Lockman Hole	&	646.2	$\pm$	1.6	&	873.8	$\pm$	2.8	&	1137.2	$\pm$	5.9	&	1519.1	$\pm$	6.0	\\
7	&	104556.84+570747.0	&	SWIRE$-$Lockman Hole	&	440.0	$\pm$	2.5	&	537.7	$\pm$	2.9	&	676.6	$\pm$	8.2	&	902.7	$\pm$	6.7	\\
8	&	104625.02+584839.1	&	SWIRE$-$Lockman Hole	&	634.7	$\pm$	2.6	&	908.1	$\pm$	3.5	&	1285.2	$\pm$	10.0	&	1688.9	$\pm$	7.0	\\
9	&	104633.70+571530.4	&	SWIRE$-$Lockman Hole	&	553.6	$\pm$	2.8	&	672.1	$\pm$	3.5	&	792.0	$\pm$	9.1	&	905.2	$\pm$	6.6	\\
10	&	104840.28+563635.6	&	SWIRE$-$Lockman Hole	&	383.3	$\pm$	2.3	&	561.0	$\pm$	3.0	&	812.9	$\pm$	9.7	&	1054.0	$\pm$	6.9	\\
11	&	104857.92+560112.3	&	SWIRE$-$Lockman Hole	&	231.6	$\pm$	2.1	&	291.2	$\pm$	2.0	&	370.1	$\pm$	7.7	&	595.5	$\pm$	6.3	\\
12	&	105000.21+581904.2	&	SWIRE$-$Lockman Hole	&	921.3	$\pm$	4.2	&	1229.6	$\pm$	5.0	&	1731.7	$\pm$	13.7	&	2337.9	$\pm$	10.0	\\
13	&	105106.12+591625.1	&	SWIRE$-$Lockman Hole	&	971.1	$\pm$	4.0	&	1210.5	$\pm$	4.0	&	1641.8	$\pm$	12.8	&	2068.6	$\pm$	7.9	\\
14	&	105518.08+570423.5	&	SWIRE$-$Lockman Hole	&	408.0	$\pm$	2.8	&	543.4	$\pm$	3.5	&	791.5	$\pm$	10.4	&	1174.7	$\pm$	8.6	\\
15	&	105604.00+581523.4	&	SWIRE$-$Lockman Hole	&	437.1	$\pm$	2.9	&	612.5	$\pm$	3.0	&	924.1	$\pm$	10.9	&	1326.3	$\pm$	7.4	\\
16	&	105959.93+574848.1	&	SWIRE$-$Lockman Hole	&	1189.6	$\pm$	4.6	&	1526.6	$\pm$	5.7	&	1993.4	$\pm$	15.5	&	2708.3	$\pm$	11.5	\\
17	&	160015.68+552259.9	&	SWIRE$-$ELAIS N1	&	841.3	$\pm$	2.7	&	1163.9	$\pm$	4.8	&	1621.5	$\pm$	9.4	&	2040.1	$\pm$	9.5	\\
18	&	160128.54+544521.3	&	SWIRE$-$ELAIS N1	&	1133.8	$\pm$	4.3	&	1854.3	$\pm$	5.3	&	2751.8	$\pm$	16.4	&	3983.0	$\pm$	11.2	\\
19	&	160341.44+541501.5	&	SWIRE$-$ELAIS N1	&	328.9	$\pm$	2.0	&	391.1	$\pm$	2.8	&	489.3	$\pm$	7.8	&	589.4	$\pm$	6.3	\\
20	&	160523.10+545613.3	&	SWIRE$-$ELAIS N1	&	682.4	$\pm$	3.6	&	982.4	$\pm$	4.7	&	1411.2	$\pm$	12.8	&	1787.4	$\pm$	9.6	\\
21	&	160630.60+542007.5	&	SWIRE$-$ELAIS N1	&	735.7	$\pm$	2.7	&	1077.8	$\pm$	3.5	&	1599.1	$\pm$	10.5	&	2302.4	$\pm$	7.5	\\
22	&	160908.95+533153.2	&	SWIRE$-$ELAIS N1	&	634.1	$\pm$	2.5	&	831.9	$\pm$	3.1	&	1136.3	$\pm$	8.3	&	1451.7	$\pm$	6.4	\\
23	&	163031.46+410145.6	&	SWIRE$-$ELAIS N2	&	625.3	$\pm$	3.3	&	713.3	$\pm$	3.7	&	874.3	$\pm$	9.9	&	1036.0	$\pm$	7.1	\\
24	&	163135.46+405756.4	&	SWIRE$-$ELAIS N2	&	582.9	$\pm$	2.6	&	789.0	$\pm$	3.1	&	1090.2	$\pm$	9.7	&	1457.1	$\pm$	6.5	\\
25	&	163143.76+404735.6	&	SWIRE$-$ELAIS N2	&	682.6	$\pm$	3.4	&	808.4	$\pm$	3.9	&	1001.6	$\pm$	9.4	&	1320.2	$\pm$	7.7	\\
26	&	163352.34+402115.5	&	SWIRE$-$ELAIS N2	&	460.8	$\pm$	2.9	&	578.2	$\pm$	3.4	&	797.3	$\pm$	9.7	&	1083.8	$\pm$	7.3	\\
27	&	163502.80+412952.9	&	SWIRE$-$ELAIS N2	&	901.4	$\pm$	3.2	&	1186.5	$\pm$	4.9	&	1521.2	$\pm$	10.6	&	1961.0	$\pm$	9.2	\\
28	&	163854.62+415419.5	&	SWIRE$-$ELAIS N2	&	532.0	$\pm$	2.3	&	673.3	$\pm$	3.2	&	850.5	$\pm$	7.9	&	1131.0	$\pm$	6.7	\\
29	&	171126.94+585544.2	&	xFLS	&	641.4	$\pm$	65.1	&	813.2	$\pm$	82.4	&	951.0	$\pm$	101.2	&	1216.0	$\pm$	124.6	\\
30	&	171334.02+595028.3	&	xFLS	&	763.2	$\pm$	76.8	&	1015.4	$\pm$	102.3	&	1456.1	$\pm$	148.7	&	1732.3	$\pm$	175.2	\\
31	&	171736.90+593011.4	&	xFLS	&	457.5	$\pm$	46.7	&	613.3	$\pm$	62.0	&	771.2	$\pm$	84.2	&	1366.4	$\pm$	138.8	\\
32	&	171748.43+594820.6	&	xFLS	&	647.8	$\pm$	65.3	&	898.1	$\pm$	90.9	&	1198.5	$\pm$	124.3	&	1570.8	$\pm$	160.1	\\
33	&	171818.14+584905.2	&	xFLS	&	675.0	$\pm$	68.3	&	869.1	$\pm$	88.1	&	1152.5	$\pm$	122.7	&	1493.2	$\pm$	152.2	\\
34	&	172104.75+592451.4	&	xFLS	&	362.1	$\pm$	36.9	&	453.4	$\pm$	46.6	&	642.1	$\pm$	70.0	&	853.8	$\pm$	89.2	\\
35	&	172414.05+593644.0	&	xFLS	&	591.2	$\pm$	59.5	&	751.8	$\pm$	76.0	&	853.0	$\pm$	89.0	&	1124.8	$\pm$	114.8	\\
\hline				
\end{tabular}      
\label{table:controlirac}             
\end{table*}

\section{Observations and Data Reduction}\label{observations}

\subsection{LoBALs}

\subsubsection{Spitzer IRS} \label{sec:irsobs}

Mid-infrared spectra of 20 of the 22 LoBALs were obtained with the $Spitzer$ Infrared Spectrograph \citep[IRS;][]{Houck2004} as part of our Cycle 5 GO program (Program ID 50792). The IRS observations of the last two scheduled objects, SDSS J023102$-$083141 and SDSS J023153$-$093333, could not be completed due to the depletion of the cryogen and the early commencement of the $Warm$ $Spitzer$ $mission$.  All 20 of the objects were observed in staring mode with the Short Low first-order module (SL1), which covers 7.4$-$14.5 $\mu$m, and with the Long Low second-order module (LL2), which covers 14.0$-$21.3 $\mu$m.  Observations in SL1 and LL2, respectively, consisted of a 6 or 14 s ramp for 2 cycles with two nod positions and a 120 or 30 s ramp for one to three cycles with two nod positions. The exact ramp durations and number of cycles for each object are listed in Table~\ref{table:lobalobs}. The slit width was 3{\farcs}7 and 10{\farcs}5 (corresponding to 24 kpc and 67 kpc at $z$=0.55), for the SL1 and LL2 orders, respectively. 

\begin{figure*}[htb]
\epsscale{1.0}
\plotone{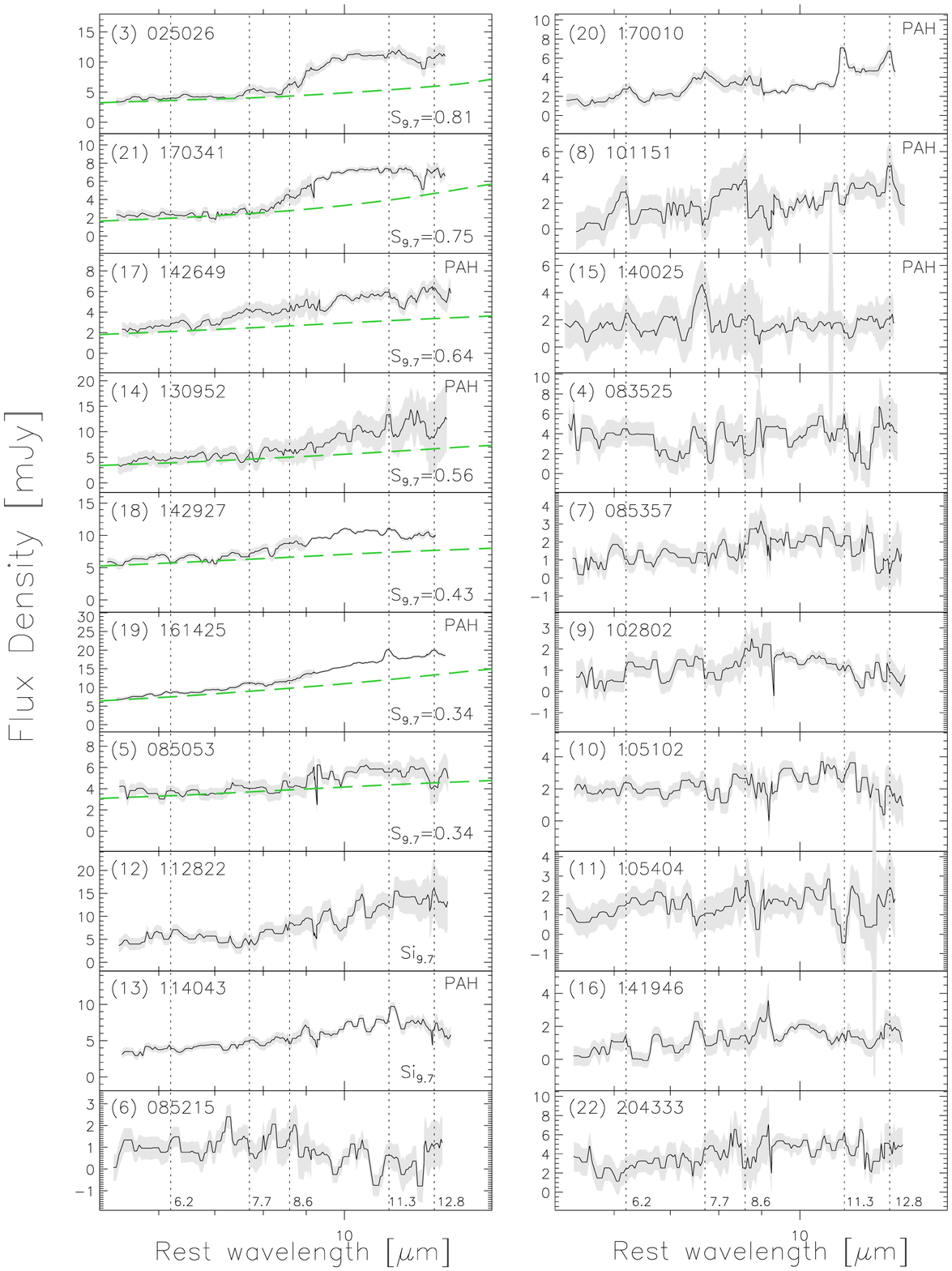}
\caption{\footnotesize Low-resolution IRS spectra of the LoBALs, plotted as flux density in units of mJy vs. rest-frame wavelength in $\mu$m. The spectra are in solid black, with 1$\sigma$ errors in gray. Plotted with a long-dash green line is the spline-interpolated continuum used to estimate the silicate strength. Vertical dotted lines at 6.2, 7.7, 8.6, 11.3, and 12.8 $\mu$m indicate the fiducial peaks of prominent PAH features. The objects in the left column show silicate emission at 9.7 $\mu$m, listed in order of decreasing strength from top to bottom. Note that 114043 and 085053 show apparent redshifted silicate emission, the strength of which could not be measure because the line is truncated and MIPS photometry was not available to constrain the SED in the FIR. Also note that 085215 shows possible silicate absorption. The top three objects in the right columns show the strongest PAH features. All objects labeled 'PAH' in the top right corner show at least one of the PAH lines. The displayed spectra have been median smoothed with a boxcar of five.}
\label{fig:lobalirs}
\end{figure*}

We used the pipeline coadded, non-subtracted post-BCD frames. The bad pixels in the images were removed with the  interactive IDL procedure IRSCLEAN1.9 (version 1.7). Sky subtraction was achieved by subtracting the two-dimensional image at one nod position from the other nod position of the corresponding order. One-dimensional spectra were extracted with the Java-based $Spitzer$ IRS Custom Extraction software (SPICE; version 2.3 Final) using the default parameters of the optimal extraction method for point sources. We used IDL to combine the two nod position in each order and then to combine the spectra from the two orders into a continuous spectrum spanning the wavelength range of 7.4$-$21.3 $\mu$m: first, the spectrum at each nod position was median-smoothed and interpolated on a uniform wavelength grid; then, the two nod positions for each order were averaged; and finally, the two orders (e.g., SL1 and LL2) were concatenated by averaging the spectra in the overlapping region. The combined mid-IR spectra of the 20 LoBALs observed with the IRS are shown in Fig.~\ref{fig:lobalirs}.

\begin{table*}
\footnotesize
\caption{LoBALs: $Sptizer$ MIPS and IRS observing log.}
\centering
\begin{tabular}{ccc|c|c|c|c|c|c|c|c|}
\hline \hline
\#	&	SDSS Object ID	&	z	&	MIPS	&	\multicolumn{7}{c}{IRS}													\\
	&		&		&		&		&	\multicolumn{3}{c|}{SL1}					&	\multicolumn{3}{c|}{LL2}					\\
	&		&		&	AORKEY	&	AORKEY	&	ramp	&	\# of 	&	integration	&	ramp	&	\# of 	&	integration	\\
	&		&		&		&		&	 duration (s)	&	cycles	&	time (s)	&	 duration (s)	&	cycles	&	time (s)	\\
\hline\hline																					
1	&	J023102.49$-$083141.2	&	0.596	&	26914048	&	$\cdots$	&	$\cdots$	&	$\cdots$	&	$\cdots$	&	$\cdots$	&	$\cdots$	&	$\cdots$	\\
2	&	J023153.63$-$093333.5	&	0.587	&	26912256	&	$\cdots$	&	$\cdots$	&	$\cdots$	&	$\cdots$	&	$\cdots$	&	$\cdots$	&	$\cdots$	\\
3	&	J025026.66+000903.4	&	0.554	&	26914560	&	26920192	&	14	&	1	&	29	&	30	&	3	&	189	\\
4	&	J083525.98+435211.2	&	0.568	&	$\cdots$	&	26918656	&	6	&	1	&	13	&	14	&	1	&	29	\\
5	&	J085053.12+445122.5	&	0.541	&	26910208	&	26915840	&	14	&	1	&	29	&	30	&	3	&	189	\\
6	&	J085215.66+492040.8	&	0.566	&	26912768	&	26918400	&	14	&	1	&	29	&	120	&	1	&	244	\\
7	&	J085357.87+463350.6	&	0.550	&	26911744	&	26917376	&	14	&	1	&	29	&	30	&	3	&	189	\\
8	&	J101151.95+542942.7	&	0.536	&	26909952	&	26915584	&	6	&	2	&	25	&	30	&	2	&	126	\\
9	&	J102802.32+592906.6	&	0.535	&	$\cdots$	&	26915328	&	14	&	1	&	29	&	120	&	1	&	244	\\
10	&	J105102.77+525049.8	&	0.543	&	26910464	&	26916096	&	14	&	1	&	29	&	30	&	3	&	189	\\
11	&	J105404.73+042939.3	&	0.578	&	$\cdots$	&	26919168	&	14	&	2	&	59	&	120	&	1	&	244	\\
12	&	J112822.41+482309.9	&	0.543	&	$\cdots$	&	26916352	&	6	&	1	&	13	&	6	&	1	&	13	\\
13	&	J114043.62+532439.0	&	0.530	&	$\cdots$	&	26914816	&	14	&	1	&	29	&	30	&	3	&	189	\\
14	&	J130952.89+011950.6	&	0.547	&	26911488	&	26917120	&	6	&	1	&	13	&	6	&	1	&	13	\\
15	&	J140025.53$-$012957.0	&	0.584	&	26913792	&	26919424	&	6	&	2	&	25	&	120	&	1	&	244	\\
16	&	J141946.36+463424.3	&	0.546	&	26911232	&	26916864	&	14	&	1	&	29	&	120	&	1	&	244	\\
17	&	J142649.24+032517.7	&	0.530	&	28968448	&	26915072	&	14	&	2	&	59	&	120	&	2	&	488	\\
18	&	J142927.28+523849.5	&	0.594	&	26914304	&	26919936	&	14	&	2	&	59	&	120	&	2	&	488	\\
19	&	J161425.17+375210.7	&	0.553	&	26912000	&	26917632	&	14	&	2	&	59	&	120	&	2	&	488	\\
20	&	J170010.83+395545.8	&	0.577	&	26913280	&	26918912	&	14	&	1	&	29	&	120	&	1	&	244	\\
21	&	J170341.82+383944.7	&	0.554	&	26912512	&	26918144	&	14	&	1	&	29	&	120	&	1	&	244	\\
22	&	J204333.20$-$001104.2	&	0.545	&	26910976	&	26916608	&	6	&	1	&	13	&	30	&	1	&	63	\\
\hline 
\end{tabular}
\label{table:lobalobs}
\end{table*}

\subsubsection{Spitzer MIPS} 

Far infrared photometry at 24, 70, and 160 $\mu$m was obtained with the Multiband Imaging Photometer for $Spitzer$ \citep[MIPS;][]{Rieke2004} as part of our GO program (ID 50792). Although the entire sample of 22 LoBALs was scheduled for MIPS observations, only 17 targets could be observed prior to the cryogen depletion and the commencement of the $Warm$ $Spitzer$ $mission$. The 24  and 70 $\mu$m observations were obtained using the small-field default resolution photometry mode, while for the 160 $\mu$m observations we used the small-field enhanced resolution mode. Typical observing modes were one 3 s cycle at 24 $\mu$m (48 s integration time), one 10 s cycle at 70 $\mu$m (126 s integration time), and four 10 s cycles at 160 $\mu$m (179 s integration time). The standard tasks of cosmic-ray removal, image-coaddition, and dark subtraction were carried out by the automated MIPS data reduction pipeline at the $Spitzer$ Science Center. Our data reduction started from the pipeline BCD files, which were assessed to be of sufficient quality. For the 24 $\mu$m observations, final mosaic images were constructed with the MOsaicker and Point source EXtractor software \citep[MOPEX:][]{Makovoz2005} after flat-fielding and background correction. At 70 and 160 $\mu$m, we used the pipeline-filtered BCDs to construct mosaics for most of the sources. Aperture photometry at 24, 70, and 160 $\mu$m was performed with IDL, using an aperture radius of 13$\arcsec$, 35$\arcsec$, and 48$\arcsec$, respectively. The MIPS fluxes are listed in Table~\ref{table:ir}.  All of the 17 sources observed with MIPS were detected at 24 microns, but only four were detected at 70 $\mu$m and only three at 160 $\mu$m. Small field photometry with MIPS 160 $\mu$m, in particular, has the problem that the filtering step of the data reduction leads to a loss of flux (Sajina 2012, in preparation). The manual states this loss is 10\% \citep[see also][]{Sajina2008}, but it can be as high as 30$-$50\%.  Three sigma upper limits at 70 $\mu$m and 160 $\mu$m were estimated from the standard deviation images associated with the mosaics. All fluxes have been corrected for the finite aperture size by multiplying by a correction factor of 1.16 at 24 $\mu$m, 1.22 at 70 $\mu$m, and 1.601 at 160 $\mu$m (aperture corrections from the MIPS Data handbook).  The systematic uncertainties are 4\% at 24 $\mu$m, 5\% at 70 $\mu$m, and 12\% at 160 $\mu$m. The uncertainties listed in Table~\ref{table:ir} for the LoBAL sample are from the aperture photometry calculation.

\subsubsection{Near-Infrared and Optical photometry: SDSS and 2MASS} \label{sdss2mass}

All of the 22 LoBALs in the sample have available SDSS Data Release Seven (DR7) and Two Micron All Sky Survey \citep[2MASS;][]{Skrutskie2006} photometry published in the quasar catalog of \citet{Schneider2010}. Table~\ref{table:sdss} lists the best SDSS PSF $ugriz$ AB magnitudes \citep[ugriz;][]{Fukugita1996}, corrected for Galactic extinction using the map of \citet{Schlegel1998}.  We impose more conservative limits on the near-infrared data than \citet{Schneider2010} by considered reliable only 2MASS magnitudes with Photometry Quality Flag (ph\_qual) A, B, or C and Read Flag (rd\_flg) 1, 2 or 3, which ensure measurements with signal-to-noise greater than five and measurement uncertainty less than 0.2.  The meaning of the flags can be found in \citet{Cutri2003}.  Table~\ref{table:ir} lists the 2MASS J (1.25 $\mu$m), H (1.65 $\mu$m), and K$_s$ (2.16 $\mu$m) Vega magnitudes of the LoBALs.

\subsection{Control sample}

\subsubsection{SDSS, 2MASS, $Spitzer$ IRAC and MIPS photometry}

We use the SDSS DR7 u, g, r, i, z and 2MASS J, H, K$_s$ photometry published by \citet{Schneider2010} together with the IRAC 3.6, 4.5, 5.8, 8.0  $\mu$m, and the MIPS 24 $\mu$m data published by \citet{Richards2006}. All the $Spitzer$ data are taken from the xFLS and SWIRE ELAIS-N1, ELAIS-N2, and Lockman Hole areas.  To obtain 70 and 160 $\mu$m fluxes, we performed aperture photometry with IDL on the processed mosaics provided online by \citet{Frayer2006} for the xFLS field\footnote{xFLS data from \\$<$http://data.spitzer.caltech.edu/popular/fls/extragalactic\_FLS/\\enhanced\_MIPS\_Ge/images/$>$} and by \citet{Lonsdale2003} for the three SWIRE fields\footnote{SWIRE data from\\ $<$http://swire.ipac.caltech.edu/swire/astronomers/data\_access.html$>$}.  The 3$\sigma$ upper limits at 70 and 160 $\mu$m were estimated from the uncertainty images. Table~\ref{table:controlirac} lists the fields from which the MIPS photometry for individual objects was extracted.\\

\section{Analysis and Results}

\subsection{MIR spectral features of LoBALs}

The combined SL1 and LL2 orders covers an observed wavelength range $\lambda_{obs} = 7.4 - 21.3$ $\mu$m, which at redshift 0.5$-$0.6 translates to rest-frame $\lambda_{rest} \approx  5-13$ $\mu$m, after removing the noisy region on the red side.  LoBALs exhibit a wide range of mid-IR spectral properties (Fig.~\ref{fig:lobalirs}). About one third (40\%) of the spectra are featureless low-signal-to-noise continua, nearly half (45\%) of the objects show silicate emission near 10 $\mu$m and a quarter (25\%) have at least one of the PAH emission lines, with concurrent PAH and silicate emission present in one tenth (10\%) of the sample. 

\subsubsection{PAH emission features}

The emission from polycyclic aromatic hydrocarbon molecules (PAHs), producing prominent lines in the mid-IR peaking at 3.3, 6.2, 7.7, 8.6, 11.3, and 12.8 $\mu$m \citep{Gillett1973, Leger1984}, is powered by moderate UV radiation.  PAHs are observed in photo-dissociation regions (PDRs) where young bright stars are contiguous with dense molecular clouds, conditions found in star-forming regions and reflection nebulae \citep[\eg][]{Duley1991, Verstraete1996}. The correlation between PAH emission and star formation rates in star-forming galaxies \citep[\eg][]{Genzel1998, Roussel2000, Dale2002} established PAHs as tracers of star formation (but see \citet{Haas2002} for a counter argument). On one hand, there is evidence that the PAH molecules are destroyed by the extreme UV and X-rays radiation \citep{Puget1989, Voit1992, Allain1996, Genzel1998, Smith2007, ODowd2009, Hunt2010}, so PAH-derived star formation may underestimate the actual activity. In AGN, PAH features are absent \citep[\eg][]{Roche1991, LeFloch2001}, weak \citep[\eg][]{Laurent2000}, or have low equivalent widths \citep[\eg][]{Clavel2000}, implying destruction or inability of the nuclear radiation to excite the aromatic feature. On the other hand, the AGN may enhance the PAH emission since the nuclear continuum contributes ample flux in the UV, which may modify the grain distribution and directly excite the PAH emission \citep[\eg][]{Smith2007}. Thus, adopting PAH emission as a star formation tracer in AGN hosts should be done with caution. 

Studies of the spatial distribution of the PAH emission in nearby AGN find that the aromatic emission arises in an extended circum-nuclear region \citep[\eg][$<$ 1 kpc]{Cutri1984} or in the galactic disk \citep{Laurent2000}, which together with the low equivalent widths of PAH features \citep[\eg][]{Roche1991, Clavel2000} implies that PAHs in AGN are predominantly excited by star-formation \citep[\eg][and references therein]{Shi2007}.  This is also supported by \citet{Schweitzer2006} who find the same ratio between the 7.7 $\mu$m PAH and the 60 $\mu$m luminosity in a sample of 27 PG QSOs and in starburst-dominated ULIRGs, suggesting that the starburst is producing all of the QSO FIR emission. 

In our sample of 20 LoBALs with available IRS spectra, we detect weak PAH features only in five of the objects. The PAH emission is strongest for the three objects shown at the top of the right column in Fig.~\ref{fig:lobalirs}. The entire complex of PAH lines is seen only in SDSS J170010+395545, while PAHs are concurrent with silicate emission only in two sources. We confirm the previously observed low incidence of PAH emission in AGN, in general, to be true also for LoBALs, in particular. 

\begin{figure}[htpb]
\plotone{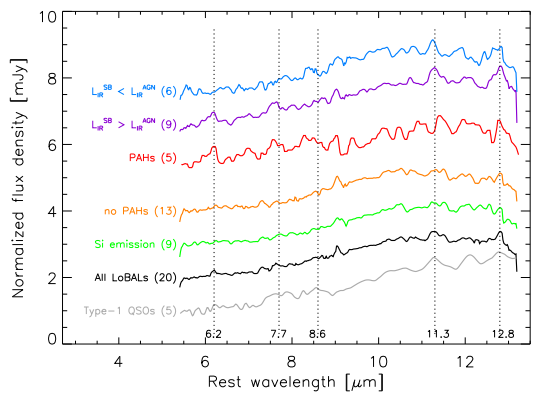}
\epsscale{1.0}
\caption{Signal-to-noise-weighted average spectra of the LoBALs in our sample, grouped according to shared characteristics. The average spectrum of all LoBALs ({\bf solid black line}) was obtained by averaging the individual spectra after normalizing their flux density by the average flux density from 5$-$6 $\mu$m. With a \textcolor{orange}{solid orange} line is the average spectrum of the objects which show no or very weak PAH features. In \textcolor{red}{solid red} is the average of the LoBALs with PAH features. In \textcolor{green}{solid green} we plot the average of the LoBALs with silicate emission at 10 $\mu$m. In \textcolor{purple}{solid purple} is the average of the LoBALs with starburst luminosity higher than the AGN luminosity from 8$-$1000 $\mu$m, while in \textcolor{blue}{solid blue} with starburst luminosity less than the AGN luminosity. For comparison, in \textcolor{light-gray}{solid gray} at the bottom we also plot the average spectrum of the five type-1 QSOs in the \citet{Hiner2009} sample which fall within the redshift range $0.5<z<0.63$. Vertical dotted lines at 6.2, 7.7, 8.6, 11.3, and 12.8 $\mu$m indicate the position of prominent PAH features. The PAH lines are stronger in the composite of objects with $L_{IR}^{SB} > L_{IR}^{AGN}$ than in the "PAH average," suggesting that the FIR emission, indeed, arises from star formation rather than from a very extended torus, for example. 
}
\label{fig:avglobal}
\end{figure}

The average MIR spectrum of all LoBALs (Fig.~\ref{fig:avglobal}), obtained by averaging the signal-to-noise-weighted individual spectra after normalization to the 6 $\mu$m continuum flux, shows very weak PAH emission at 6.2, 11.3 and 12.8 $\mu$m. Plotted in the same figure are the average spectra of LoBALs grouped according to shared MIR spectral characteristics, i.e., objects with PAH emission, those with silicate emission, and those showing neither silicate nor PAHs. PAH emission at 6.2 $\mu$m becomes more prominent in the average spectrum of the LoBALs with PAHs. In the average spectrum of LoBALs which did not otherwise show individual PAH features, we detect the 12.8 $\mu$m line. We note that the feature at 12.8 $\mu$m might be a blend of the PAH at 12.7 $\mu$m and the low-excitation fine-structure emission line of [\ion{Ne}{2}] 12.8 $\mu$m, which is dominant in \ion{H}{2} regions and used as SF tracer. In QSOs, however, [\ion{Ne}{2}] could also arise in the narrow line region of the AGN and \citet{Veilleux2009} show that the starburst contributes at most 50\% to its flux. In addition, we subdivided the 15 LoBALs that had both IRS spectra and FIR constraints into those with AGN infrared luminosity greater than and less than their starburst luminosity (see section~\ref{measuredquantities} for details on the measurement of these quantities). The two averages are shown in Fig.~\ref{fig:avglobal}, denoted $L_{IR}^{SB} < L_{IR}^{AGN}$ and $L_{IR}^{SB} > L_{IR}^{AGN}$, respectively. We note that in the cases of FIR upper limits for objects with $L_{IR}^{SB} > L_{IR}^{AGN}$, it is not certain whether $L_{IR}^{SB}$ is less than or greater than $L_{IR}^{AGN}$.

The PAH lines are stronger in the composite of objects with $L_{IR}^{SB} > L_{IR}^{AGN}$ than in the "PAH average," suggesting that the FIR emission, indeed, arises from star formation rather than from a very extended torus, for example. The average of all LoBALs is remarkably similar to the subgroups of those without PAHs and the one with silicate emission. For comparison, we also plot an average spectrum of the five type-1 QSOs in the \citet{Hiner2009} sample which fall within the redshift range $0.5 < z < 0.63$. The composite of all LoBALs is similar to that of their type-1 QSOs.

\subsubsection{Silicate emission feature at 10 $\mu$m}

Silicate compounds comprise the majority of the interstellar dust and manifest themselves in the MIR via two main features centered around 10 $\mu$m and 18 $\mu$m. The wavelength coverage of our IRS spectra allows us to investigate the silicates peaking at 9.7 $\mu$m which arise from the stretching of the Si$-$O bond \citep{Knacke1973}. Due to low dust temperatures of order $T \sim 100$ K, this feature is also seen to peak at longer wavelengths, up to $\lambda_{Si} \sim$ 11 $\mu$m \citep[\eg][]{Zakamska2008}. The exact peak of the feature is speculated to be sensitive to grain size and composition, with larger dust grains and crystalline dust (as opposed to amorphous dust) causing the feature to peak at longer wavelengths \citep[\eg][]{Bouwman2001}. However, \citet{Nikutta2009} show that the flat-topped silicate emission peaks observed in several type-1 sources can be explained by simple radiative transfer effects in standard Galactic dust in a clumpy torus model.

Most studies of silicate detections in AGN show that there is a notable dependence of the silicate strength on the optical classification of the AGN. In the framework of the unification model, silicate emission arising from the dust torus surrounding the accretion disk was predicted by the models of \citet{Pier1993}. If we attribute the appearance of the silicate features in AGN to viewing angle, then type-1 AGN are expected to show silicate in emission and type-2 objects in absorption. In fact, most observations of AGN in which the feature is detected support this interpretation. Type 2 quasars \citep[\eg][]{Zakamska2008, Lacy2007, Hiner2009} and Seyfert 2  galaxies \citep{Hao2007} are almost exclusively characterized by silicate absorption, while both weak silicate emission and absorption are found in type-1 QSOs  \citep[\eg][]{Siebenmorgen2005, LHao2005, Schweitzer2008, Hiner2009, Landt2010} and Seyfert 1 galaxies \citep{Hao2007}. 

However, silicate emission or absorption is not ubiquitously present in AGN. The occasional detection of silicate emission in type-2 AGN  \citep[\eg][]{Sturm2005, LHao2005, Teplitz2006, Lacy2007, Hiner2009} and silicate absorption in type-1 objects  \citep[\eg][]{Weedman2005, Hao2007}, as well as the absence of the silicate feature in many AGN, challenges the universality of the orientation model and suggests a more complicated scenario. For instance, with the exception of two objects, one mini-LoBAL and one red QSO, all of the type-1 and type-2 QSOs in the \citet{Hao2007} sample exclusively show weak silicatein emission which argues against orientation-dependent silicate emission in QSOs, at least.   Models of clumpy torus geometries \citep{Nenkova2002, Nenkova2008a, Nenkova2008b} and/or larger dust grain sizes \citep[\eg][]{Laor1993, Maiolino2001} offer a solution to these discrepancies, but can only reproduce weak silicate features.  \citet{Nikutta2009}, for instance, show that clumpy dust geometry of the obscuring region can explain both why the 10 $\mu$m feature is not seen in deep absorption in any AGN and why it has been detected in emission in type-2 sources.

In order to compare our LoBALs to other studies, we calculate the silicate strength at 9.7 $\mu$m, $S_{9.7}$, as:

\begin{center}
$S_{9.7}$ = ln $\frac{f_{peak}(9.7 \mu m )}{f_{cont}(9.7 \mu m)}$,
\end{center}

\noindent
where $f_{peak}(9.7 \mu m)$ is the observed flux density at the peak of the silicate feature and $f_{cont}(9.7 \mu m)$ is the continuum flux density interpolated below the peak of the emission line. The common challenge of determining the underlying continuum is further exacerbated by the limited wavelength range of our IRS spectra (i.e., rest-frame 5.0$-$14.0 $\mu$m). As is apparent in Fig.~\ref{fig:lobalirs}, some of the silicate features have a truncated red wing and are often redshifted. Fortunately, the MIPS photometry allows us to extend the range of the IRS spectra by modeling the SEDs, which includes the silicate emission amplitude as a free parameter in the fit (see Section \ref{sedmodeling}).  Using the overall SED model, we determine the continuum by following the fitting recipes of \citet{Spoon2007} for continuum-dominated sources and interpolate the local mid-IR continuum over the range 5.0$-$31.5 $\mu$m by fitting a cubic spline to the 5.0$-$7.5, 14.0$-$14.2, and 26.1$-$31.5 $\mu$m continuum regions. 

In our sample of LoBALs, silicate is present exclusively in emission and is detected in nine of the 20 objects, ranging from S$_{9.7}$ = 0.34$-$0.81. Hence, LoBALs exhibit the weak silicate emission typical of other type-1 QSOs \citep[\eg][]{Siebenmorgen2005, LHao2005, Haas2005, Cleary2007, Hiner2009}. However, we note that the detection/non-detection of silicate dust emission might be a function of the signal-to noise ratio of the data (see Section~\ref{trends}). The strength of the feature is individually noted in the bottom right corner of each spectrum in Fig.~\ref{fig:lobalirs}. The interpolated continuum is over-plotted with a dashed green line. Although apparently present, the strength of the feature could not be estimated in two objects, SDSS J112822+482308 and SDSS J114043+532439, due to lack of MIPS photometry at long wavelengths and, consequently, an SED model which allows better determination of the continuum. We note that SDSS J085215+492040 shows possible broad, weak absorption, but the IRS spectrum for that object has very low signal-to-noise ratio and our inability to measure the feature precludes us from further speculation. Silicate absorption features is never observed to peak at wavelengths longer than 9.8 $\mu$m \citep[\eg][]{Nikutta2009}, while the peak of this apparent dip is located at $\sim$11 $\mu$m.

\subsection{SED modeling} \label{sedmodeling}

In general, the infrared emission in galaxies hosting an AGN is a combination of (1) starburst emission from dust heated by the UV flux of O and B stars in active star-forming regions \citep{Devereux1990, Devereux1997}, (2) the AGN emission from the dusty torus reprocessing the accretion disk continuum and re-radiating it in the infrared \citep{Pier1993}, (3) diffuse, ambient cold dust (cirrus) emission illuminated by the interstellar radiation field \citep[\eg][]{Rowan-Robinson1989}, and (4) infrared emission from evolved stellar populations \citep{Knapp1992, Mazzei1994}. 

We fit the optical-through-FIR SEDs of the LoBALs and the control sample of non-LoBALs with a multi-component empirical model, which allows us to disentangle the two major sources that power the FIR emission, i.e., starburst and AGN. In our phenomenological SED modeling approach, we assume that cirrus and evolved stellar populations have negligible contribution to the FIR power budget, and account only for the starburst and AGN contributions.

The available optical photometry of our LoBAL sample is listed in Table~\ref{table:sdss}, the MIPS photometry in Table \ref{table:ir}, and the IRS spectroscopy is plotted in Fig. \ref{fig:lobalirs}. The photometry of the control sample is given in Tables~\ref{table:sdss} ,~\ref{table:ir}, and~\ref{table:controlirac}. The fitting code is described in detail by \citet{Sajina2006} and \citet{Hiner2009}. Here we briefly outline the modeling procedure. The SEDs of LoBALs are fit by four components: 

(1) a QSO component, constructed from the line-free continuum of the \citet{Richards2006} SED composite and the emission lines from the SDSS quasar composite of \citet{VandenBerk2001} (this modification was necessary to reduce the host galaxy contribution at long wavelengths present in the \citet{VandenBerk2001} composite);

(2) a hot mid-IR component, modeled as a power law with an exponential cutoff at short and long wavelengths, with a turndown at $\sim$ 20 $\mu$m; 

(3) a warm FIR component, accounting for the small grain dust emission spanning a wide range of temperatures is modeled as a power law with cutoffs at high and low frequencies;

(4) and a cold FIR component, modeled as a modified black body at a fixed temperature $T_{dust} \sim$ 45 K.  Constraining the dust temperature to 45 K is a conservative assumption for a typical ULIRG-level starburst in nuclear regions, which are probably the closest local analogues to starbursts in powerful quasars.  

(5) The SEDs of the control sample require an additional near-infrared (NIR) component accounting for the emission from very hot dust, which was modeled as a modified black body at a temperature of $T$=1000 K. This very hot dust component, which is thought to be emission from the inner wall of the dust torus, is not necessary for fitting the SEDs of the LoBALs, with the possible exception of SDSS J105102+525049 and SDSS J170010+395554. Since we do not have NIR data for the majority of the LoBALs of reliable quality (see section $\S$\ref{sdss2mass}), most of their SEDs are not well constrained in that region and we cannot say for certain that the 3 $\mu$m bump is not present in LoBALs.

The model also includes the Small Magellanic Cloud (SMC) extinction law \citep{Prevot1984, Bouchet1985} and a composite PAH template or 9.7 $\mu$m silicate emission, if applicable.  

The FIR emission from cold and warm dust is attributed to star formation, while the mid-IR hot dust is assumed to be heated by the AGN.  For the objects not detected in the FIR, 3$\sigma$ fluxes were used to impose upper limits to the SEDs.

In Fig.~\ref{fig:lobalseds} we show the SEDs of the 15 LoBALs for which both IRS spectra and MIPS photometry could be obtained. The SEDs of the control sample are plotted in Fig.~\ref{fig:controlfits}. Overlaid are the best-fit SED model and the phenomenological break-up of the different components. Although we have observed 20 of the 22 LoBALs with the $Spitzer$ IRS (see Fig.~\ref{fig:lobalirs}), five of those could not be observed with MIPS due to the early cryogen depletion.

\begin{figure*}[htpb]
\epsscale{1.0}
\plotone{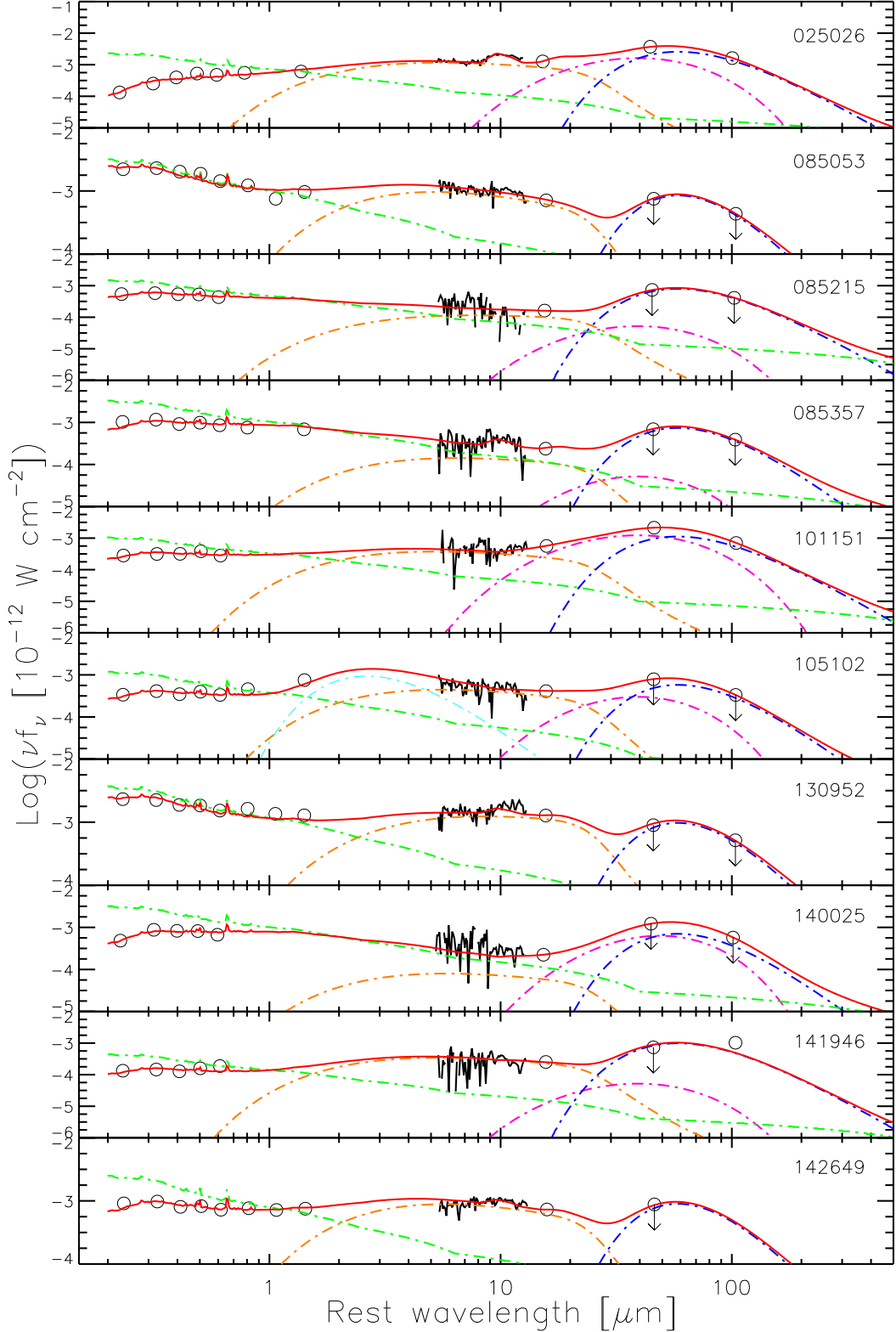}
\caption{\footnotesize Model fits to the SEDs of the 15 LoBALs in the sample with available $Spitzer$ IRS  and MIPS observations. The solid black lines are the $Spitzer$ IRS spectra and the open black circles are SDSS $ugriz$ and $Spitzer$ MIPS photometry at 24, 70, and 160 $\mu$m. The SED model is described in detail in \S \ref{sedmodeling}. The \textcolor{red}{overall fit} to the SED is plotted with a solid red line. The individual components to the fit are plotted with dot-dash lines, color-coded as follows: \textcolor{green}{unreddened SDSS quasar composite} in green; \textcolor{cyan}{ near-IR modified black body at temperature $T$=1,000 K} in cyan; \textcolor{orange}{modified mid-IR power-law} component in orange; \textcolor{magenta}{warm small grain dust} in magenta; and \textcolor{blue}{45K modified black body} component in blue. Down arrows indicate 3$\sigma$ upper limits. Partial object names are indicated in the upper right corner of each plot.}
\label{fig:lobalseds}
\end{figure*}

\addtocounter{figure}{-1}
\begin{figure*}[htpb]
\epsscale{1.0}
\plotone{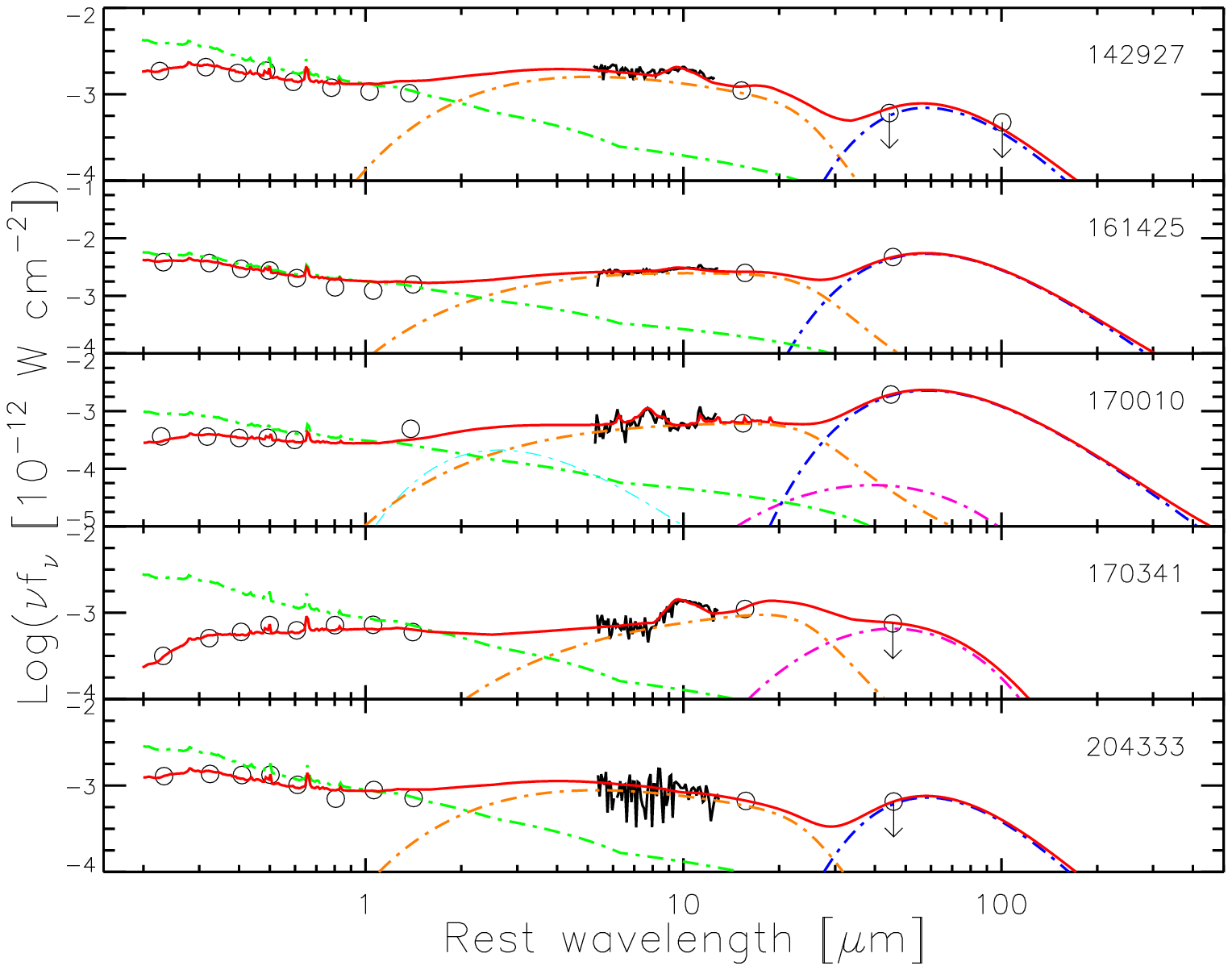} 
\caption{Continued.}
\end{figure*}

\begin{figure*}[htpb]
\epsscale{1.0}
\plotone{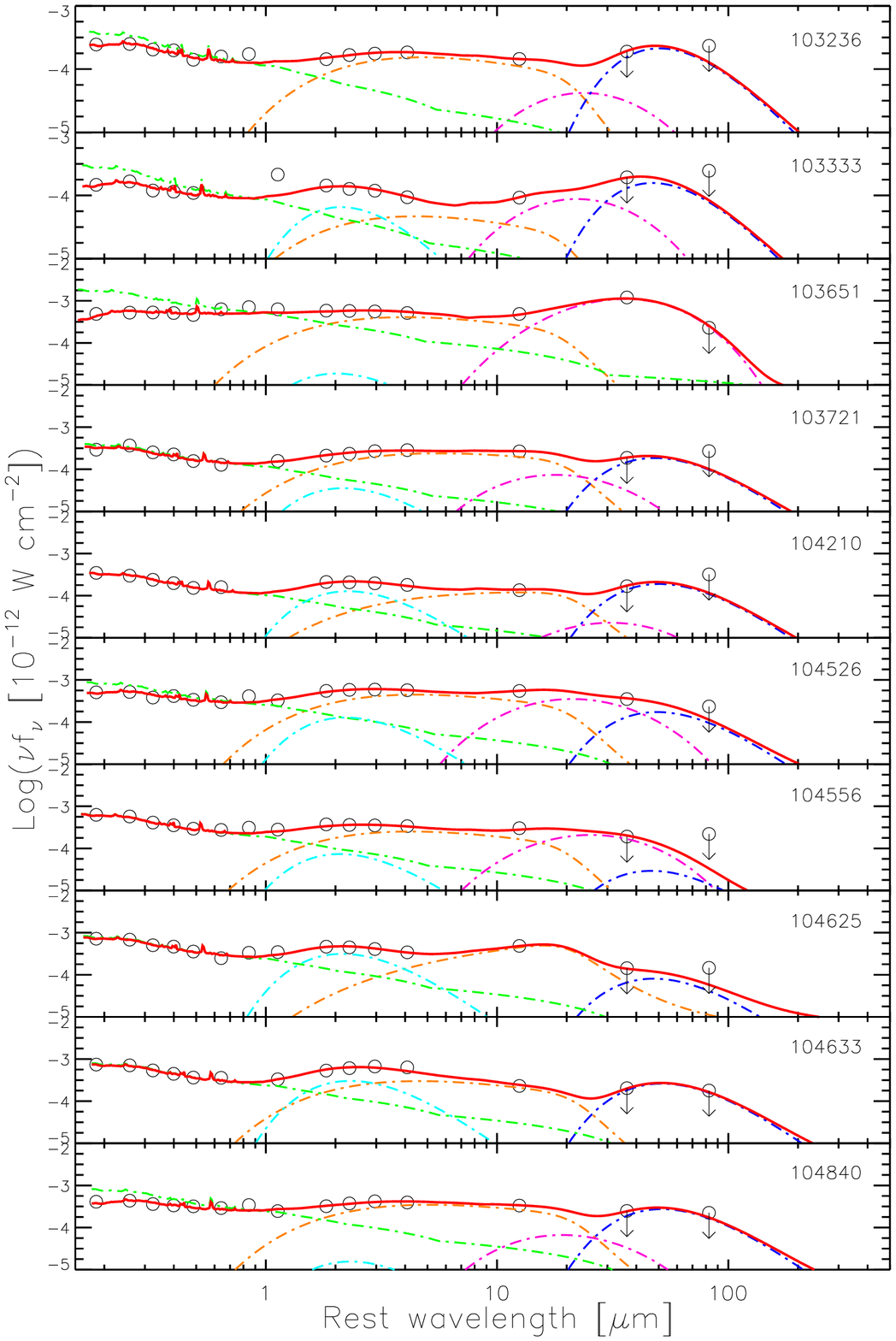}
\caption{\footnotesize SED models of the control sample of 35 non-LoBALs.  The solid black lines are the $Spitzer$ IRS spectra and the open black circles are SDSS $ugriz$, $Spitzer$  IRAC 3.6, 4.5, 5.8, 8.0  $\mu$m, and MIPS photometry at 24, 70, and 160 $\mu$m. The \textcolor{red}{overall fit} to the SED is plotted with a solid red line. The individual components to the fit are plotted with dot-dash lines, color-coded as follows: \textcolor{green}{unreddened SDSS quasar composite} in green; \textcolor{cyan}{ near-IR modified black body at temperature $T$=1,000 K} in cyan; \textcolor{orange}{modified mid-IR power-law} component in orange; \textcolor{magenta}{warm small grain dust} in magenta; and \textcolor{blue}{45K modified black body} component in blue. Down arrows indicate 3$\sigma$ upper limits. Partial object names are indicated in the upper right corner of each plot, with the exception of the additional very hot dust component, plotted as dot-dashed cyan line.}
\label{fig:controlfits}
\end{figure*}

\addtocounter{figure}{-1}
\begin{figure*}[htpb]
\epsscale{1.0}
\plotone{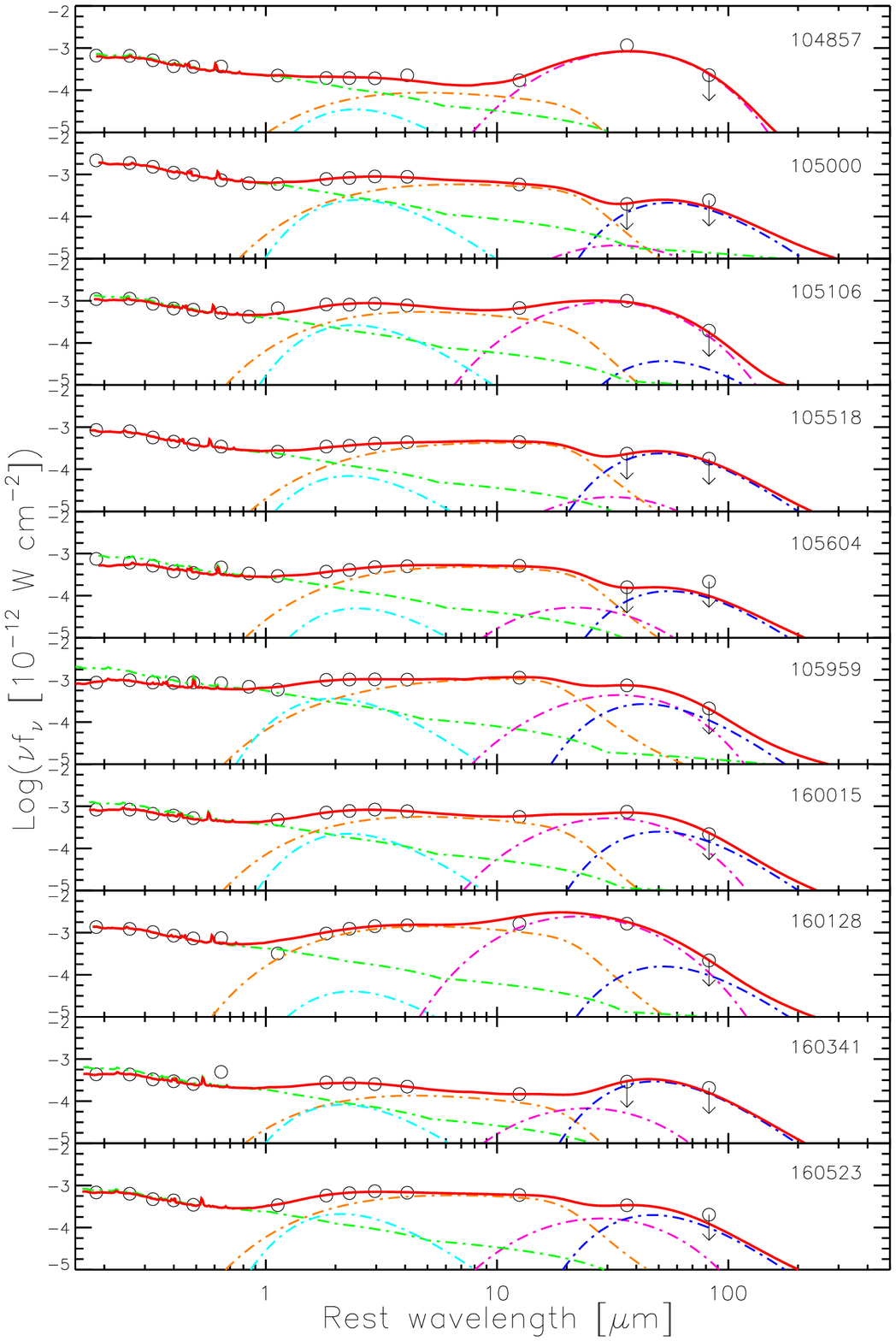}
\caption{Continued.}
\end{figure*}

\addtocounter{figure}{-1}
\begin{figure*}[htpb]
\epsscale{1.0}
\plotone{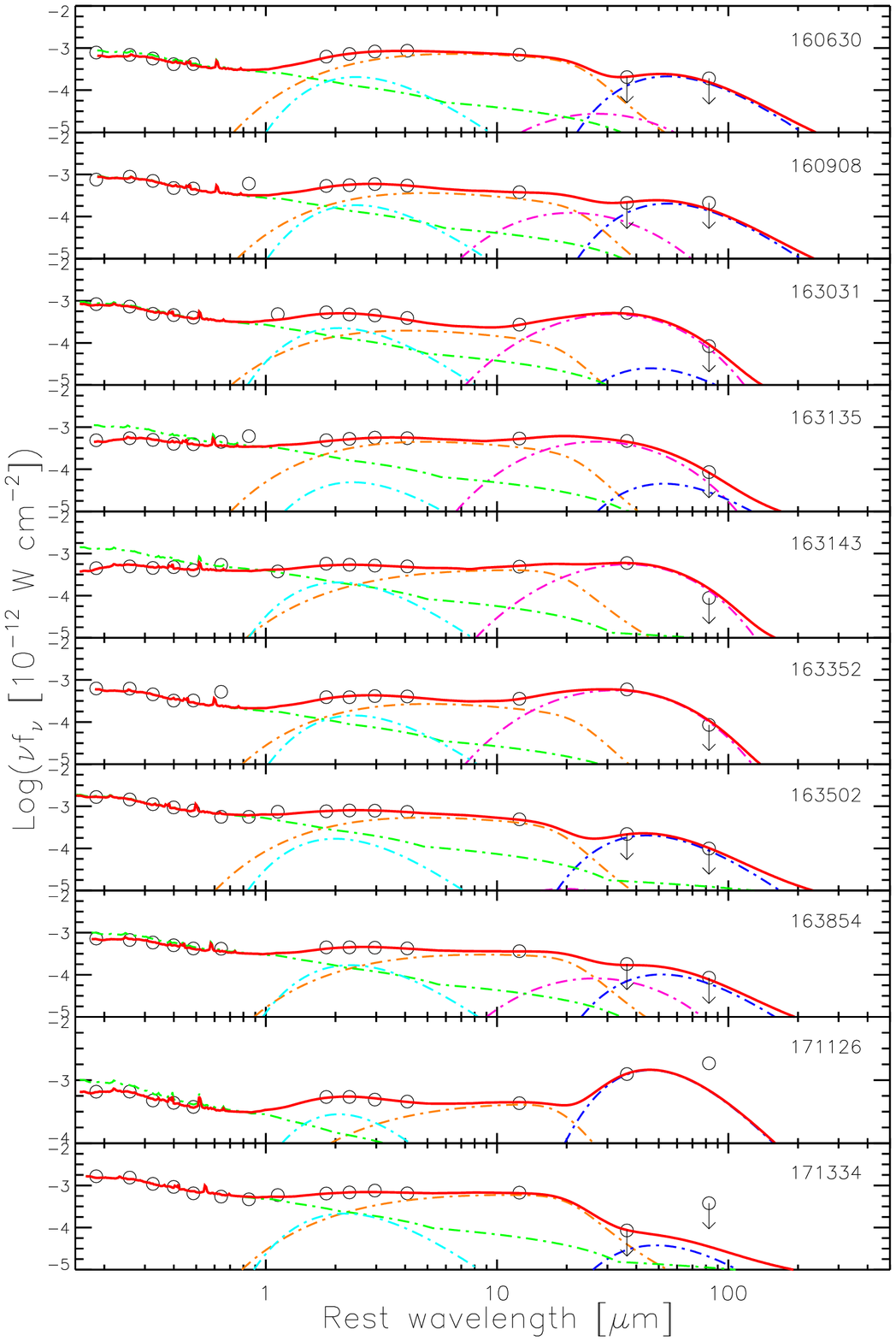}
\caption{Continued.}
\end{figure*}

\addtocounter{figure}{-1}
\begin{figure*}[htpb]
\epsscale{1.0}
\plotone{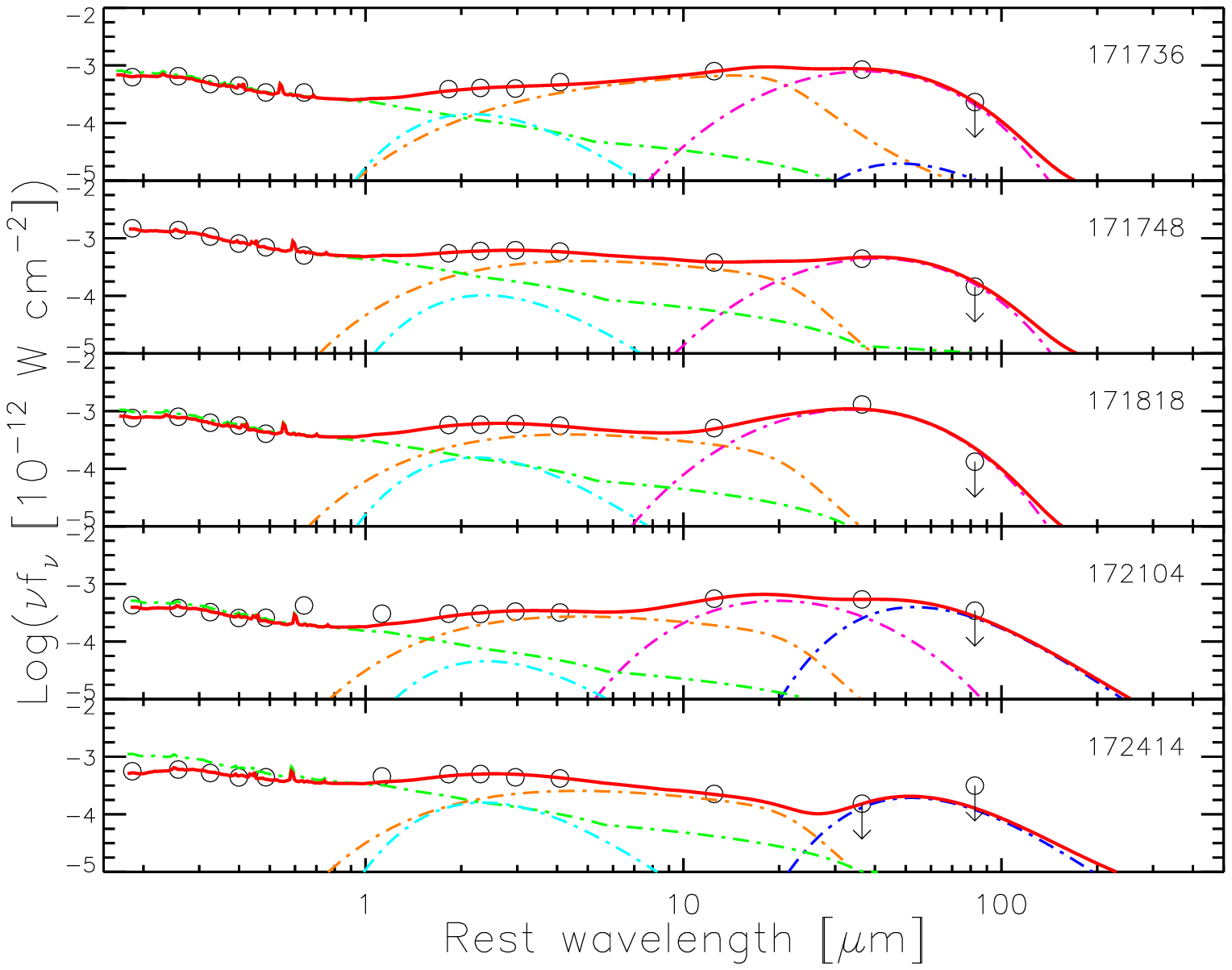}
\caption{Continued.}
\end{figure*}

\subsection{Measured quantities}  \label{measuredquantities}

Our phenomenological SED modeling is used to break down the AGN and starburst contributions to the FIR emission. In order to quantify the infrared luminosities and the level of star formation activity in LoBALs and compare it to that in non-LoBALs, we estimate the following quantities, which we list in Table~\ref{table:loballum} and in Table~\ref{table:controllum} for the samples of LoBALs and non-LoBALs, respectively. In these tables, columns (1) and (2) list the object number in the sample and its name.

Column(3) gives the total infrared luminosity from 8 to 1000 $\mu$m, $L_{IR,8-1000\mu m}^{total}$, which is integrated from the best-fit SED model and includes the contribution of the starburst and the AGN to the infrared flux.

Column (4) lists  the FIR luminosity contributed by the starburst, $L_{IR}^{SB}$. It is estimated by summing the warm and cold components of the model. In our SED model, those two components account for the emission from warm, small grain dust and cold dust at $T_{dust} \sim$ 45 K, respectively.  We tested the robustness of the choice of a 45 K dust by allowing $T_{dust}$ to vary as a free parameter in the fit, and found that it introduces a variation in the estimate of $L_{IR}^{SB}$ of less than 5\%.  Attributing the warm and cold FIR emission to star formation is empirically justified by \citet{Lacy2007} who find, for a sample for six type-2 AGN, that the sum of these components scales with PAH luminosity for a wide range of FIR luminosities. Additional evidence for the star formation origin of the FIR emission comes from \citet{Netzer2007} who find that to be the case for PG QSOs.

Column (5) lists  the contribution of the AGN (hot dust + continuum) to the FIR luminosity, $L_{IR,8-1000\mu m}^{AGN}$. It was estimated by integrating the reddened hot dust component and the QSO continuum from 8 to 1000 $\mu$m. 

Column (6) shows the relative percentage contribution of the starburst infrared luminosity to the total IR luminosity from 8 to 1000 $\mu$m, $ \frac{L_{IR}^{SB}}{L_{IR,8-1000\mu m}^{total}}\times 100\%$, obtained as the ratio of the integrated QSO composite and hot dust from 8 to 1000 $\mu$m to the total integrated flux from 8 to 1000 $\mu$m.

In Column (7) we list the total infrared luminosity from 3$-$1000 $\mu$m, $L_{TIR}^{MIPS} $. While we note that this quantity is calibrated for galaxies, we estimate it solely for comparison with other studies. It is calculated from its empirically calibrated relationship to the broad-band MIPS fluxes at 24, 70, and 160 $\mu$m derived by \citet[eq. 4]{Dale2002}:

\begin{center}
$L_{TIR}^{MIPS} = \zeta_1 \nu L_{\nu}(24\mu m)+\zeta_2 \nu L_{\nu}(70\mu m)+\zeta_3 \nu L_{\nu}(160\mu m)$
\end{center}

\noindent
where [$\zeta_1, \zeta_2, \zeta_3$] are redshift dependent coefficients (e.g., [3.91499, 0.48179, 1.0049] for $z$ = 0.55). The authors state that this relationship reproduces the model bolometric infrared luminosities from 3 to 1000 $\mu$m to better than 1\% for galaxies at z = 0, less than 4\% for all other redshifts $z<4$, and 7\% for colder galaxies. The data used to constrain the Dale \& Helou SED models ranges in L$_{TIR}$ from less than 10$^8$ L$_{\odot}$ to 10$^{12}$ L$_{\odot}$, hence, this method should be successful in estimating L$_{TIR}$ for our sample of ULIRGs and LIRGs. Five of the LoBALs in our sample were not observed with MIPS. For three of those we used Scanpi 2\footnote{Scanpi 2 is the scan processing and integration tool for extraction of IRAS photometry, developed at IPAC/Infrared Science Archive, which is operated by the Jet Propulsion Laboratory, California Institute of Technology, under contract with the National Aeronautics and Space Administration.} to extract IRAS photometry at 25, 60, and 100 $\mu$m from which we estimate their total IR luminosity, following \citet[eq. 5]{Dale2002}.  The upper limits of L$_{TIR}$ are calculated using the detections and/or the 3$\sigma$ flux values for the bands in which the source was not detected. We note that the IRAS fluxes may not be reliable due to the large beam size and possible contamination by neighboring infrared sources. For instance, we measure the IRAS flux at 25 $\mu$m of SDSS J161425+375210 with Scanpi 2 to be 130 mJy, while its MIPS flux at 24 $\mu$m is 20 mJy. 

Column (8) gives the SFRs calculated from the starburst infrared luminosities listed in columns (4). The SFRs are estimated using the \citet{Kennicutt1998} relationship:

\begin{center}
 $SFR(M_{\odot} yr^{-1}) = 4.5 \times 10^{-44} L_{IR}$ 
\end{center}

\noindent
where L$_{IR}$ is total infrared luminosity in erg/s, within the range 8$-$1000 $\mu$m.

Note that the \citet{Kennicutt1998} SFR relationship uses a slightly different definition of L$_{TIR}$ from the one derived with the Dale \& Helou (2002) formulae, 8$-$1000 $\mu$m as opposed to 3$-$1000 $\mu$m. The starburst components of the SEDs do not extend blueward of 7 $\mu$m in any object, so this does not affect our SFR$_{IR}^{SB}$.

For the five objects with IRS spectra that do not have MIPS photometry, we fit the SDSS photometry and the IRS data with an AGN composite and mid-IR hot dust component only, and, thus, estimate only the AGN contribution to the FIR. Without FIR data, it is impossible to estimate the starburst contribution to the FIR flux in those cases. However, modeling allows us to estimate lower limits to $L_{IR,8-1000\mu m}^{total}$ from the AGN flux from 8$-$1000 $\mu$m. For all other quantities, upper limits indicate that 3$\sigma$ MIPS fluxes were used to constrain the SED model in the FIR.

\begin{table*}
\begin{center}
\caption{LoBALs: Infrared luminosities and SFRs.}
\begin{tabular}{ccrrrcrc}
\hline \hline
\#	&	SDSS Object ID	&		$Log(L_{IR,8-1000\mu m}^{total})$		&		$Log(L_{IR}^{SB})$ 	&	$Log(L_{IR,8-1000\mu m}^{AGN})$ 	&	$\frac{L_{IR}^{SB}}{L_{IR,8-1000\mu m}^{total}}$	&		$Log(L_{TIR}^{MIPS})$ 		&		$SFR_{IR}^{SB}$						\\
	&		&		[$Log(L_{\odot})$]		&		[$Log(L_{\odot})$]	&	[$Log(L_{\odot})$]	&	[\%]	&		[$Log(L_{\odot})$]		&		[$M_{\odot}$ yr$^{-1}$]						\\
(1)	&	(2)	&		(3)		&		(4)	&	(5)	&	(6)	&		(7)		&		(8)						\\
\hline\hline																										
1	&	J023102.49$-$083141.2	&		$\cdots$		&		$\cdots$	&	$\cdots$	&	$\cdots$	&	$<$	12.08		&		$\cdots$						\\
2	&	J023153.63$-$093333.5	&		$\cdots$		&		$\cdots$	&	$\cdots$	&	$\cdots$	&	$<$	11.63		&		$\cdots$						\\
3	&	J025026.66+000903.4	&		{\bf 12.39}		&		12.25	&	11.81	&	73	&		{\bf 12.49}	&		310	$_{-	30	 } ^{+	25	}$	\\
4	&	J083525.98+435211.2	&	$>$	11.61		&		$\cdots$	&	$\cdots$	&	$\cdots$	&	13.47	$^\star$		&		$\cdots$						\\
5	&	J085053.12+445122.5	&	$<$	11.87		&	$<$	11.46	&	11.65	&	38	&	$<$	12.01		&	$<$	49						\\
6	&	J085215.66+492040.8	&	$<$	11.64		&	$<$	11.52	&	11.03	&	75	&	$<$	11.65		&	$<$	56						\\
7	&	J085357.87+463350.6	&	$<$	11.64		&	$<$	11.45	&	11.20	&	63	&	$<$	11.69		&	$<$	48						\\
8	&	J101151.95+542942.7	&		{\bf 12.01}		&		11.93	&	11.23	&	83	&		{\bf 12.04}		&		148	$_{-	23	 } ^{+	16	}$	\\
9	&	J102802.32+592906.6	&	$>$	11.04		&		$\cdots$	&	$\cdots$	&	$\cdots$	&		$\cdots$		&		$\cdots$						\\
10	&	J105102.77+525049.8	&	$<$	11.72		&	$<$	11.49	&	11.32	&	59	&	$<$	11.83		&	$<$	53						\\
11	&	J105404.73+042939.3	&	$>$	11.28		&		$\cdots$	&	$\cdots$	&	$\cdots$	&		$\cdots$		&		$\cdots$						\\
12	&	J112822.41+482309.9	&	$>$	11.98		&		$\cdots$	&	$\cdots$	&	$\cdots$	&	13.39	$^\star$		&		$\cdots$						\\
13	&	J114043.62+532439.0	&	$>$	11.77		&		$\cdots$	&	$\cdots$	&	$\cdots$	&	13.47	$^\star$		&		$\cdots$						\\
14	&	J130952.89+011950.6	&	$<$	12.03		&	$<$	11.56	&	11.85	&	33	&	$<$	12.24		&	$<$	62						\\
15	&	J140025.53$-$012957.0	&	$<$	11.86		&	$<$	11.76	&	11.17	&	79	&	$<$	11.85		&	$<$	98						\\
16	&	J141946.36+463424.3	&		11.71		&		11.56	&	11.18	&	70	&		11.84		&		63						\\
17	&	J142649.24+032517.7	&	$<$	11.85		&	$<$	11.49	&	11.60	&	43	&	$<$	11.95		&	$<$	52						\\
18	&	J142927.28+523849.5	&	$<$	12.07		&	$<$	11.48	&	11.94	&	25	&	$<$	12.27		&	$<$	52						\\
19	&	J161425.17+375210.7	&		{\bf 12.52} 	&		12.28	&	12.15	&	57	&		{\bf 12.56}		&		326	$_{-	19	 } ^{+	20	}$	\\
20	&	J170010.83+395545.8	&		{\bf 12.12}		&		11.97	&	11.59	&	70	&		{\bf 12.06}		&		161	$_{-	18	 } ^{+	20	}$	\\
21	&	J170341.82+383944.7	&	$<$	11.90		&	$<$	11.42	&	11.72	&	33	&	$<$	12.15		&	$<$	45						\\
22	&	J204333.20$-$001104.2	&	$<$	11.84		&	$<$	11.40	&	11.64	&	36	&	$<$	11.92		&	$<$	43						\\	
\hline
\end{tabular}
\vspace{-0.1cm}
\begin{flushleft}
{\footnotesize
Notes: Bold face denotes a {\bf ULIRG}, i.e., $L_{IR} > 10^{12} L_{\odot}$ \\ 
$^\star$ Indicates $L_{TIR}$ estimated from IRAS rather than MIPS fluxes.
}
\end{flushleft}
\label{table:loballum}       
\end{center}
\end{table*}

\begin{table*}
\caption{Control sample of non-LoBALs: Infrared luminosities and SFRs.}
\centering
\begin{tabular}{ccrrcccr}
\hline\hline																			
\#	&	SDSS Object ID	&		$Log(L_{IR}^{8-1000\mu m})$	&		$Log(L_{IR}^{SB})$ 	&	$Log(L_{IR,8-1000\mu m}^{AGN})$	&	$\frac{L_{IR}^{SB}}{L_{IR}^{8-1000 \mu m}}$	&		$Log(L_{TIR}^{MIPS})$	&		$SFR_{IR}^{SB}$	\\
	&		&		[$Log(L_{\odot})$]	&		[$Log(L_{\odot})$]	&	[$Log(L_{\odot})$]	&	[\%]	&		[$Log(L_{\odot})$]	&		[$M_{\odot}$ yr$^{-1}$]	\\
(1)	&	(2)	&		(3)	&		(4)	&	(5)	&	(6)	&		(7)	&		(8)	\\
\hline\hline																			
1	&	J103236.22+580033.9	&	$<$	11.47	&	$<$	11.20	&	11.14	&	53	&	$<$	11.69	&	$<$	27	\\
2	&	J103333.92+582818.8	&	$<$	11.12	&	$<$	11.00	&	10.49	&	76	&	$<$	11.39	&	$<$	17	\\
3	&	J103651.94+575950.9	&		11.72	&		11.55	&	11.22	&	68	&		11.86	&		61	\\
4	&	J103721.15+590755.7	&	$<$	11.44	&	$<$	11.08	&	11.19	&	43	&	$<$	11.75	&	$<$	20	\\
5	&	J104210.25+594253.5	&	$<$	11.39	&	$<$	11.10	&	11.08	&	50	&	$<$	11.68	&	$<$	21	\\
6	&	J104526.73+595422.6	&		11.81	&		11.50	&	11.51	&	49	&		12.08	&		54	\\
7	&	J104556.84+570747.0	&	$<$	11.32	&	$<$	10.97	&	11.06	&	44	&	$<$	11.69	&	$<$	16	\\
8	&	J104625.02+584839.1	&	$<$	11.50	&	$<$	10.54	&	11.44	&	11	&	$<$	11.87	&	$<$	6	\\
9	&	J104633.70+571530.4	&	$<$	11.67	&	$<$	11.25	&	11.47	&	37	&	$<$	11.85	&	$<$	30	\\
10	&	J104840.28+563635.6	&	$<$	11.77	&	$<$	11.37	&	11.55	&	39	&	$<$	12.00	&	$<$	40	\\
11	&	J104857.92+560112.3	&		{\bf 12.01}	&		11.93	&	11.20	&	84	&		12.07	&		148	\\
12	&	J105000.21+581904.2	&	$<$	12.09	&	$<$	11.36	&	12.00	&	18	&	$<$	12.36	&	$<$	39	\\
13	&	J105106.12+591625.1	&		{\bf 12.19}	&		11.96	&	11.81	&	58	&		12.38	&		158	\\
14	&	J105518.08+570423.5	&	$<$	11.80	&	$<$	11.22	&	11.66	&	26	&	$<$	12.06	&	$<$	28	\\
15	&	J105604.00+581523.4	&	$<$	11.98	&	$<$	11.26	&	11.89	&	19	&	$<$	12.31	&	$<$	31	\\
16	&	J105959.93+574848.1	&		11.72	&		11.23	&	11.55	&	32	&		11.97	&		29	\\
17	&	J160015.68+552259.9	&		{\bf 11.99}	&		11.70	&	11.68	&	51	&		12.16	&		86	\\
18	&	J160128.54+544521.3	&		{\bf 12.54}	&		12.33	&	12.13	&	61	&		12.67	&		367	\\
19	&	J160341.44+541501.5	&	$<$	11.37	&	$<$	11.18	&	10.93	&	63	&	$<$	11.53	&	$<$	25	\\
20	&	J160523.10+545613.3	&		11.69	&		11.18	&	11.53	&	30	&		11.97	&		26	\\
21	&	J160630.60+542007.5	&	$<$	12.11	&	$<$	11.35	&	12.02	&	17	&	$<$	12.41	&	$<$	38	\\
22	&	J160908.95+533153.2	&	$<$	11.92	&	$<$	11.50	&	11.70	&	38	&	$<$	12.18	&	$<$	55	\\
23	&	J163031.46+410145.6	&		11.45	&		11.27	&	10.98	&	66	&		11.60	&		32	\\
24	&	J163135.46+405756.4	&		{\bf 11.99}	&		11.65	&	11.72	&	46	&		12.22	&		77	\\
25	&	J163143.76+404735.6	&		11.65	&		11.33	&	11.36	&	48	&		11.83	&		37	\\
26	&	J163352.34+402115.5	&		{\bf 11.96}	&		11.75	&	11.53	&	62	&		12.13	&		98	\\
27	&	J163502.80+412952.9	&	$<$	11.40	&	$<$	10.72	&	11.29	&	20	&	$<$	11.64	&	$<$	9	\\
28	&	J163854.62+415419.5	&	$<$	11.70	&	$<$	11.12	&	11.56	&	26	&	$<$	11.98	&	$<$	23	\\
29	&	J171126.94+585544.2	&		11.85	&		11.68	&	11.34	&	68	&		12.10	&		82	\\
30	&	J171334.02+595028.3	&	$<$	11.69	&	$<$	10.29	&	11.67	&	3	&	$<$	12.11	&	$<$	3	\\
31	&	J171736.90+593011.4	&		11.93	&		11.61	&	11.64	&	48	&		12.17	&		71	\\
32	&	J171748.43+594820.6	&		{\bf 11.97}	&		11.62	&	11.71	&	44	&		12.12	&		72	\\
33	&	J171818.14+584905.2	&		{\bf 11.96}	&		11.80	&	11.45	&	69	&		12.10	&		109	\\
34	&	J172104.75+592451.4	&		{\bf 12.06}	&		11.92	&	11.51	&	71	&		12.33	&		144	\\
35	&	J172414.05+593644.0	&	$<$	11.65	&	$<$	11.16	&	11.48	&	32	&	$<$	11.93	&	$<$	25	\\
\hline
\end{tabular} 
\vspace{-0.1cm}
\begin{flushleft}
{\footnotesize Notes: Bold face denotes a {\bf ULIRG}. Objects with $L_{IR} > 10^{11.95} L_{\odot}$ were considered.}
\end{flushleft}     
\label{table:controllum}             
\end{table*}

\subsection{Infrared luminosities of LoBALs}

\subsubsection{Total IR luminosities from MIPS photometry}\label{sec:irlum}

First we address the total infrared luminosities of LoBALs to see if their apparent association with ULIRGs found at $z < 0.4$ \citep{Canalizo2002} is typical for all LoBALs.  We use the definition of a ULIRG by \citet{Sanders1988}, i.e., $L_{IR}>10^{12}L_\odot$.  

For fair comparison with other studies which address the total infrared luminosities based on photometry measurements alone, we use the estimates of $L_{IR}$ from MIPS, $L_{TIR}^{MIPS}$, calculated with the \citet{Dale2002} relation (Column (7) in Table~\ref{table:loballum}). 
Considering only the detections, we find that one (5\%) LoBAL is a LIRG ($L_{TIR}^{MIPS}=10^{11-12}L_\odot$) and that 7/20 (35\%) have $L_{TIR}^{MIPS}>10^{12}L_\odot$. Three of those seven ULIRGs are potential HyLIRGs ($L_{IR}>10^{13}L_\odot$) based on $IRAS$ photometry, which is unreliable due to the large beam size and possible inclusion of neighboring sources. However, we note that more than half of the objects were not detected.

Only five LoBALs in the sample have detections in the FIR MIPS bands at 70 and 160 $\mu$m. Of those five, only four are ULIRGs.  Regardless of the prevalence of upper limits, our results unambiguously show that low-redshift LoBALs are not exclusively associated with ULIRGs, with at least 40\% of them being found in LIRGs. 

Similar estimates of $L_{IR}$ are derived by integrating the flux from 8 to 1000 $\mu$m of the best-fit SED models (Fig.~\ref{fig:lobalseds}). The values of $L_{IR,8-1000\mu m}^{total}$ are listed in Table~\ref{table:loballum}, column (3). We find that 4/20 (20\%) of the LoBALs are ULIRGs and 16/20 (80\%) are LIRGs. For comparison, we estimate the total infrared luminosities, $L_{IR,8-1000\mu m}^{total}$, of the control sample of 35 non-LoBALs in the same way (see Table~\ref{table:controllum}). 
Of the 17 control QSOs with FIR detections, nine (at least 26\% of the total sample) are ULRIGs.  We find that the fraction of LoBALs residing in ULIRGs is similar to that of non-LoBALs. The median total infrared luminosities, $L_{IR,8-1000\mu m}^{total}$, of the FIR-detected subsamples are $\sim$ 12.12 for the LoBALs and 11.92 for the non-LoBALs.

 In order to compare the LoBAL and non-LoBAL samples in the presence of so many upper limits, we use the survival analysis statistical tests \citep{Feigelson1985}. We test the hypothesis that LoBALs and non-LoBALs have the same distributions of infrared luminosities. Using the Gehan and the logrank tests, we find that there is about 59\% and 24\% chance of observing the difference in medians if the LoBAL and the non-LoBAL samples were drawn from the same distribution. We conclude that LoBALs do not show statistically significant differences in infrared luminosity compared to non-LoBALs. 
 
\begin{figure}[htpb]
\epsscale{1.0}
\plotone{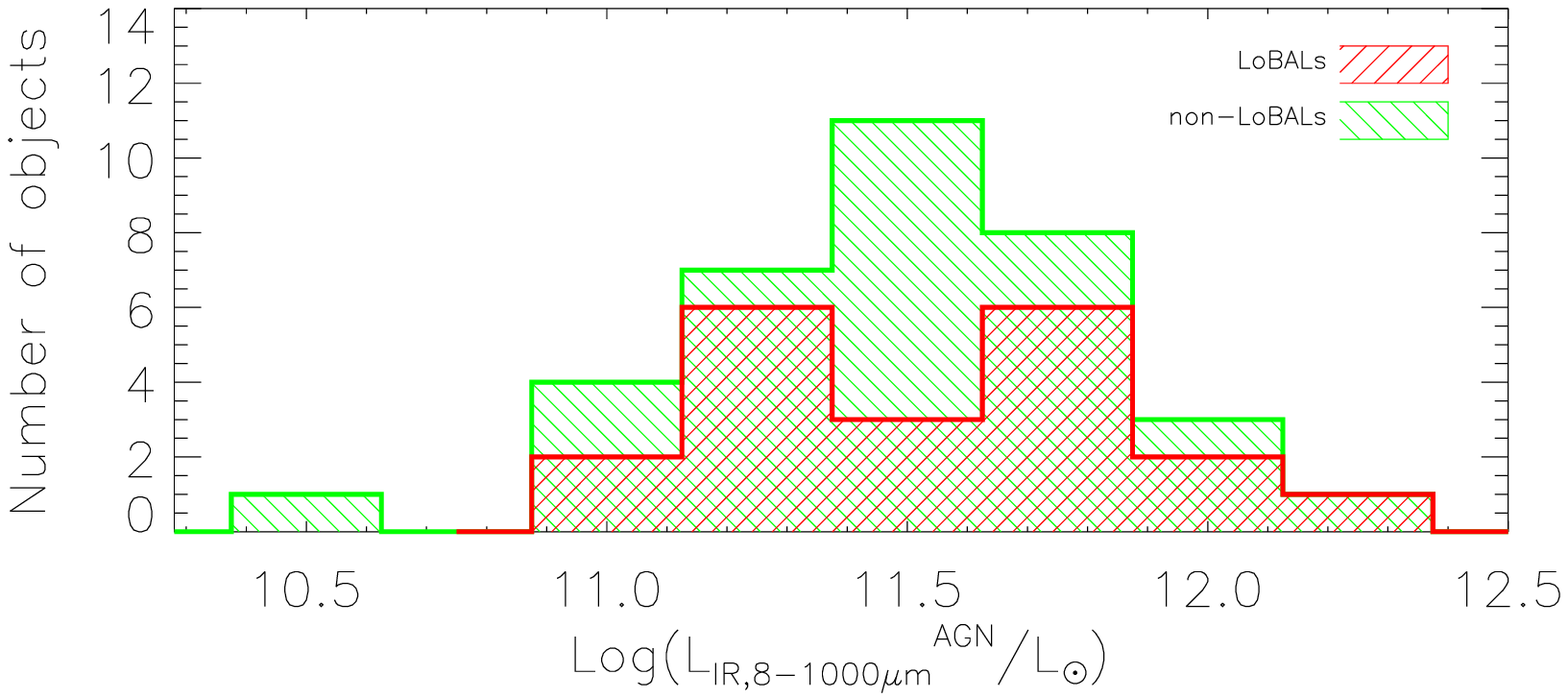}
\caption{\footnotesize LoBALs and non-LoBALs: Distribution of the AGN infrared luminosity integrated from the best-fit SED model between 8$-$ 1000 $\mu$m, plotted in bins of 0.25 dex. The medians of 11.61 for the LoBALs and 11.48 for the non-LoBALs show that LoBALs do not harbor intrinsically more infrared-luminous AGN than those in non-LoBALs.}
\label{fig:agnfirhist}
\end{figure}

 We also find that LoBALs do not harbor intrinsically more infrared luminous AGN. In Fig.~\ref{fig:agnfirhist}, we show the distributions of the AGN infrared luminosities for LoBALs and non-LoBALs. The medians for the two types of objects are 11.61 and 11.48, respectively. This implies that the infrared-to-optical ratio in LoBALs and non-LoBALs is comparable. This suggests that the dusty material obscuring the nuclear source has similar covering fractions.
 
\subsubsection{IR luminosity due to Star Formation}

Although in normal galaxies the FIR emission is usually attributed to dust emission excited by star-formation, it is now known that in AGN the central nuclear source contributes significantly to the dust heating \citep[\eg][]{Evans2006, Veilleux2009}. Hence, removing the AGN contribution to the IR emission is important when quantifying the star formation activity in the galaxy and deriving star-formation rates from the total infrared luminosity using the \citet{Kennicutt1998} relation.

We estimate the star formation contribution to the far-infrared luminosity, $L_{IR}^{SF}$, by integrating the warm and cold components of the best-fit SED model, thus, removing the AGN contribution.  The median star formation luminosity of our LoBAL sample is $log(L_{IR}^{SF}/L_{\odot}) \approx 11.52$ and $\approx 11.35$ for the control sample of non-LoBALs, when considering all objects, including those with MIPS upper limits. LoBALs have higher median star formation luminosities, $log(L_{IR,LoBAL}^{SF}/L_{\odot}) \approx 11.97$, than the non-LoBALs, $log(L_{IR,control}^{SF}/L_{\odot}) \approx 11.72$, when we compare only the subset of objects with FIR detections for which the SEDs are well constrained.

The median contribution from star formation to the total IR luminosity from 8$-$1000 $\mu$m,  $\frac{L_{IR}^{SB}}{L_{IR,8-1000\mu m}^{total}}$ (Table~\ref{table:loballum}, column (6)),  in LoBALs is 41\% and in non-LoBALs 48\% (70\% and 62\%, respectively, if only the FIR detections are considered). Although we do observe significant variations among individual objects, from as low as 25\% to as high as 83\% in LoBALs and between 3\% and 84\% in non-LoBALs, there is no significant difference between the samples. In the presence of mostly upper limits in the FIR MIPS bands, we use the Gehan and the logrank survival analysis statistical tests \citep{Feigelson1985} and find the the probability of observing the slight difference in the median values is 79\% and 96\%, respectively, if LoBALs and the control non-LoBALs were drawn from the sample parent population. 

Our estimate of the star formation activity in LoBALs is much higher than that found for all types of BALs by \citet{Gallagher2007}. They report that less than 20\% of the total FIR flux in BALs arises due to star formation. \citet{Gallagher2007} model the radio to x-ray SEDs of a large sample of 38 BALs at z $ >$ 1.4, consisting of 32 HiBALs and 6 LoBALs. Although they state that the quasar  likely dominates the far-infrared emission in BALs, they note that the two most luminous starbursts in their sample are LoBALs, and the preponderance of upper limits at far-infrared wavelengths for the majority of the Lo- and Hi-BALs hampers their ability to draw definitive conclusions on the issue.\\

\subsection{Trends with IR luminosity} \label{trends}

\begin{figure}[htbp]
\plotone{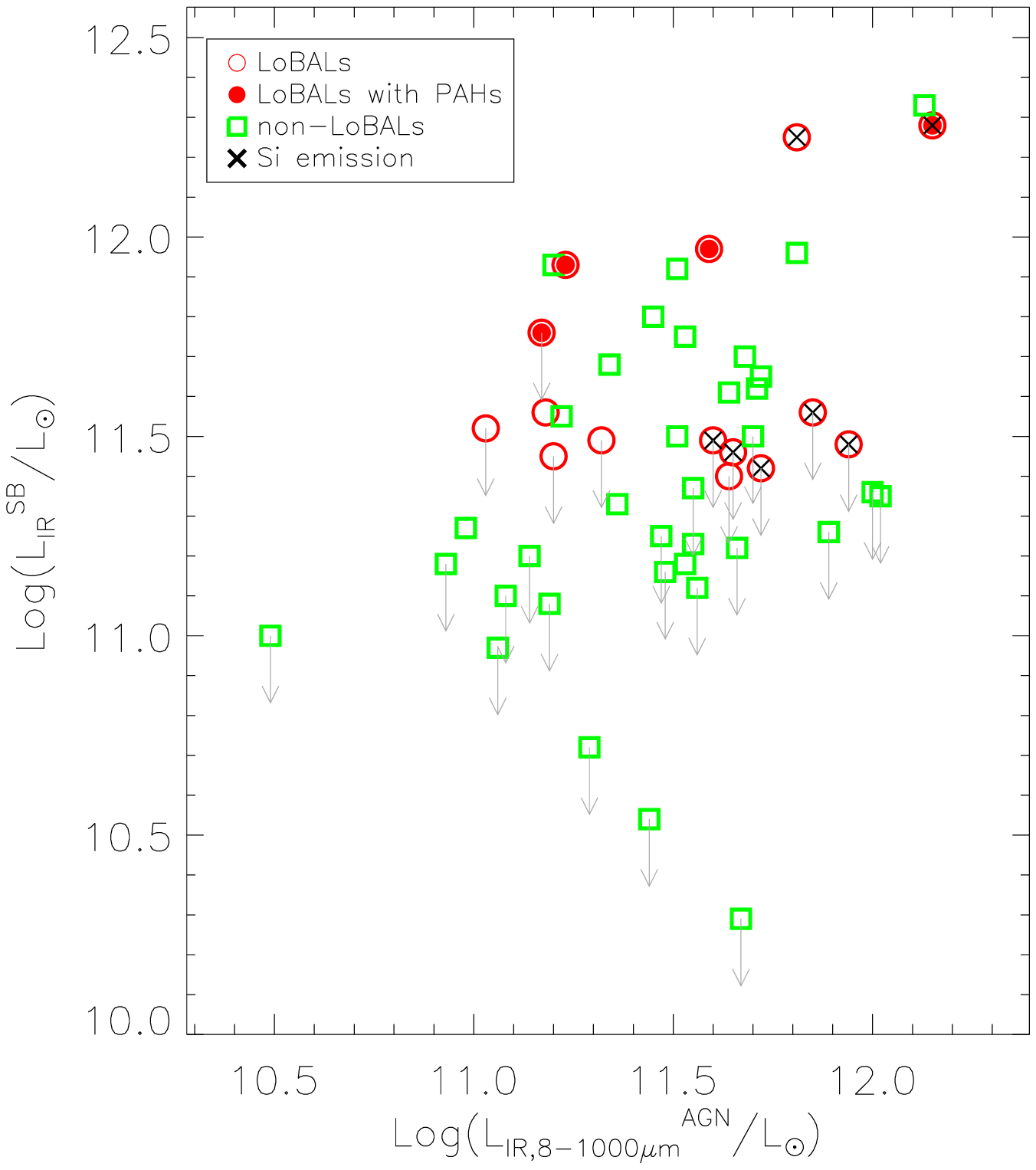}
\epsscale{0.1}
\caption{\footnotesize Starburst luminosity vs. AGN contribution to the FIR luminosity in LoBALs and in non-LoBAL QSOs, as estimated from the SED models. The data are listed in Table~\ref{table:loballum} and \ref{table:controllum}. LoBALs are plotted with \textcolor{red}{red circles}; non-LoBALs are plotted with \textcolor{green}{green squares}. Black $\times$ denotes a LoBAL with detectable 10 $\mu$m silicate emission. Filled red circles indicate the present of PAH emission, while the red open circles are LoBALs without PAH features. The presence, or absence, of silicate emission in LoBALs is correlated with the total AGN FIR luminosity, suggesting that the weak silicate emission we see in LoBALs is directly excited by the central AGN. All LoBALs with $Log(L_{IR,8-1000\mu m}^{AGN}/L_{\odot}) >$ 11.55, show silicate emission.  Although the more IR-luminous AGN appear to preferentially have silicate emission, this is likely an artifact of the systematically higher S/N of their IRS spectra. PAH emission is favored among LoBALs with starburst luminosities $Log(L_{IR}^{SB}/L_{\odot}) >$ 11.75. One exception is SDSS J025053+000903, which has the strongest silicate emission feature in our sample.}
\label{fig:agnsbfir}
\end{figure}

In Fig.~\ref{fig:agnsbfir} we plot the IR luminosity from the starburst, $L_{IR}^{SB}$, versus the AGN contribution to the IR flux from 8$-$1000 $\mu$m, $L_{IR,8-1000\mu m}^{AGN}$. At a first glance, we see that the presence of silicate emission in LoBALs is correlated with the AGN IR luminosity, which may suggest that the weak silicate emission we see in LoBALs is directly excited by the central nuclear source. All LoBALs with $log(L_{IR,8-1000\mu m}^{AGN}/L_{\odot}) >$ 11.55 show silicate emission. However, whether the silicate feature is detected is entirely dependent on the signal-to-noise ratio (S/N) of the IRS spectra. We tested this by introducing artificial noise to the spectra of the objects with detected silicate emission using the $IRAF$ task {\it mknoise}, creating spectra with S/N equal to the median of the silicate non-detections, S/N $\sim$ 1.5.  The silicate emission we observe in nine of the LoBALs would not have been detected if the IRS data had lower quality.  We conclude that the non-detection of silicate in the majority of the LoBALs might be simply a low S/N effect.

PAH emission is favored among the LoBALs with starburst IR luminosities $log(L_{IR}^{SB}/L_{\odot}) >$ 11.75. Higher total IR luminosities for objects with PAH detections are also seen by \citet{Schweitzer2006} in PG QSOs. Four of the five LoBALs with MIPS detections at 70 and 160 $\mu$m are ULIRGs (i.e., $ L_{IR,8-1000\mu m}^{total}>10^{12}L_{\odot}$). Three of those objects show the strongest PAH features: SDSS J101151+542942, SDSS J161425+375210, and SDSS J170010+395545.  There is one exception: SDSS J025026+000903 is a ULIRG and shows no PAHs, but has the strongest silicate emission feature of the entire sample.

\subsection{Star formation rates in LoBALs}

It has been shown that the total IR luminosity is a plausible star formation rate indicator (e.g., \citealt{Kennicutt1998} ($L_{IR}\sim$ 8$-$1000 $\mu$m); \citealt{Kewley2002}; \citealt{Mann2002} ($L_{IR}\sim$ 3$-$1000 $\mu$m)).  However, in AGN, much of the accretion disk UV and optical continuum is reprocessed by dust near the active nucleus and re-radiated in the infrared.   By breaking down the different components that contribute to the overall SED (i.e., AGN, hot, warm, and cold dust), we partially alleviate the problem of AGN contamination to the FIR flux.  With the explicit assumption that the starburst component of the IR luminosity is dominated by warm and cold dust re-processed light from O and B stars, we calculate SFRs from the total starburst infrared luminosity, $L_{IR}^{SB}$, using the \citet{Kennicutt1998} relationship (Table~\ref{table:loballum}, column (8)). \citet{Rieke2009} show that the choice of initial mass function (IMF) is crucial in calibrating the SFR and, by adopting a Salpeter-like slope with more shallow slope at low masses, estimate a correction to the \citet{Kennicutt1998} SFR of SFR$_{Rieke09}$ = 0.66~$\times$~SFR$_{Kennicutt98}$.  Although such an IMF fits better extragalactic star forming regions \citep{Rieke2009}, the SFRs given here are estimated with the \citet{Kennicutt1998} relationship and are not to be interpreted literally but comparatively.

Even removing the AGN contribution to the FIR flux, there is possibly still contribution from older stellar populations \citep[\eg][]{Devereux1989, Popescu2000} and from the AGN itself. \citet{Hiner2009} find evidence for non-starburst contribution to what we call the starburst IR flux. They model the SEDs of a sample of type-1 and type-2 QSOs in the same way we do, and they estimate SFRs with the \citet{Kennicutt1998} relationship using the total model IR luminosity corrected for the AGN contribution and the total integrated PAH luminosity.  They find slightly lower SFRs derived from the PAH luminosity than from the starburst IR luminosity. This discrepancy is interpreted as the presence of an additional AGN contribution to the FIR flux, which has not been accounted for in the modeling. However, we also note that PAH emission can be affected by the presence of dust because the silicate opacity curve peaks in close proximity to some of the PAH features \citep{Kemper2004}. And although mounting evidence supports the prediction that PAH carriers are destroyed by the AGN radiation \citep[\eg][]{Voit1992}, it is still not clear whether or not the presence of an AGN enhances the PAH emission because the AGN emits UV radiation, which, in principle, can excite PAH emission \citep[\eg][]{Smith2007}.

We find that the host galaxies of LoBALs have a range of star formation rates. With the caveat that most of our results are upper limits due to non-detections in the MIPS 70 and 160 $\mu$m bands, we find that the median SFR in LoBALs is on the order of $\sim$ 52 $_{\odot}$ yr$^{-1}$ and in non-LoBALs $\sim$ 38 $_{\odot}$ yr$^{-1}$.  Four LoBALs have particularly high star formation rates of $\sim$ 150 M$_{\odot}$ yr$^{-1}$, for SDSS J101151+542942 and SDSS J170010+395545, and $\sim$300 M$_{\odot}$ yr$^{-1}$, for SDSS J161425+375210 and SDSS J025026+000903. Similar fraction of the non-LoBALs have SFRs $\sim$100$-$300 M$_{\odot}$ yr$^{-1}$. We note that two of those four LoBALs (SDSS J161425+375210 and SDSS J170010+395545) and all, but one, of the highly star-forming non-LoBALs, do not have strong constrains on the FIR emission from cold dust since we only have upper limits at 160 $\mu$m. If we consider only the five LoBALs and 17 non-LoBALs with well constrained far-infrared luminosities, we find higher median star formation rate in LoBALs, $\sim$160 M$_{\odot}$ yr$^{-1}$, than in non-LoBALs, $\sim$ 90 M$_{\odot}$ yr$^{-1}$.

\section{Discussion}

\subsubsection*{Mid-IR spectral properties}

The low-resolution mid-IR spectra of LoBALs show a wide range of properties. We find that about 45\% of the LoBALs show weak 9.7 $\mu$m silicate emission (S$_{9.7}$ = 0.34$-$0.81), which is typical of other type-1 QSOs. Weak PAH emission is observed in 25\% of the LoBALs, testifying to the presence of current star-formation in their hosts. Although about 40\% of the LoBALs have featureless, low S/N spectra, we determined that the low quality of the mid-infrared data affects the detection of silicate.

The 9.7 $\mu$m silicate feature is exclusively seen in weak emission in about half of the LoBALs. This supports the previously observed dichotomy between type-1 and type-2 QSOs, that is, the silicate feature appears in weak emission in the former and in weak absorption in the latter.  Even the ULIRGs among LoBALs exhibit silicate emission rather than the typical deep silicate absorption seen in the majority of ULIRGs (\eg~\citealt{Spoon2007}; see also \citealt{CHao2005}). Hence, in terms of their silicate feature, LoBALs resemble type-1 QSOs with their typical weak silicate emission. In the context of the orientation model in which BALs are present in all type-1 QSOs, but observed only at limited viewing angles, the similarity of the silicate dust emission in LoBALs and in non-LoBALs indicates that the viewing angle for LoBALs is, not surprisingly, closer to those of type-1 QSOs than to those of type-2 QSOs.   On the other hand, in the context of an evolution model in which deeply embedded AGN evolve to become unobscured type-1 QSOs, our results would imply that LoBALs mark one of the last stages of the transition.  At this stage, the nuclear region has been cleared up of the thick dust envelope responsible for the deep absorption seen in other dust-obscured objects such as ULIRGs.

Yet, the optical spectra of our LoBALs suffer from high levels of obscuration. We estimate the AGN extinction in the optical from the SED fit, which includes SMC extinction law, and find a median value of $A_V=0.43$ and a range $A_V=0-1$ (Table~\ref{table:tau}), indicating significant levels of obscuration in some LoBALs.  LoBALs are known to have intrinsically bluer optical-UV continua than normal QSOs \citep[i.e.,][]{Richards2002}, and so the derived A$_V$ values are likely underestimated.  The non-LoBALs of the control sample suffer significantly less extinction at a median level of $A_V=0.06$ (Table~\ref{table:tau}).   We estimate the median color excess, $E(B-V)$ in LoBALs, using the SMC extinction law and $R_V$ = 2.72, to be $E(B-V) \approx$ 0.14, a value comparable to previous studies of LoBALs by \citet{Sprayberry1992} and \citet{Gibson2009}, who find SMC-type color excess of 0.12 and 0.14, respectively, but somewhat higher than the value of 0.077 reported by \citet{Reichard2003b}.

\begin{table}
\scriptsize
\caption{\small SMC-type AGN extinction in $V$ band and silicate strengths.}
\centering
\begin{tabular}{cccc}
\hline \hline
\#	&	SDSS Object ID	&	A(V)	&	$Si_{9.7}$	\\
\hline\hline
\multicolumn{4}{c}{LoBALs} \\	
\hline\hline							
1	&	J023102.49$-$083141.2	&	$\cdots$	&	$\cdots$	\\
2	&	J023153.63$-$093333.5	&	$\cdots$	&	$\cdots$	\\
3	&	J025026.66+000903.4	&	1.03	&	0.81	\\
4	&	J083525.98+435211.2	&	0.00	&	$\cdots$	\\
5	&	J085053.12+445122.5	&	0.14	&	0.34	\\
6	&	J085215.66+492040.8	&	0.41	&	$\cdots$	\\
7	&	J085357.87+463350.6	&	0.53	&	$\cdots$	\\
8	&	J101151.95+542942.7	&	0.52	&	$\cdots$	\\
9	&	J102802.32+592906.6	&	0.00	&	$\cdots$	\\
10	&	J105102.77+525049.8	&	0.50	&	$\cdots$	\\
11	&	J105404.73+042939.3	&	0.51	&	$\cdots$	\\
12	&	J112822.41+482309.9	&	0.61	&	$\cdots$	\\
13	&	J114043.62+532439.0	&	0.00	&	$\cdots$	\\
14	&	J130952.89+011950.6	&	0.14	&	0.56	\\
15	&	J140025.53$-$012957.0	&	0.69	&	$\cdots$	\\
16	&	J141946.36+463424.3	&	0.48	&	$\cdots$	\\
17	&	J142649.24+032517.7	&	0.44	&	0.64	\\
18	&	J142927.28+523849.5	&	0.28	&	0.43	\\
19	&	J161425.17+375210.7	&	0.10	&	0.34	\\
20	&	J170010.83+395545.8	&	0.42	&	$\cdots$	\\
21	&	J170341.82+383944.7	&	0.83	&	0.75	\\
22	&	J204333.20$-$001104.2	&	0.28	&	$\cdots$	\\
\hline\hline
\multicolumn{4}{c}{Control sample of non-LoBALs}\\
\hline\hline
1	&	J103236.22+580033.9	&	0.12	&	$\cdots$	\\
2	&	J103333.92+582818.8	&	0.21	&	$\cdots$	\\
3	&	J103651.94+575950.9	&	0.52	&	$\cdots$	\\
4	&	J103721.15+590755.7	&	0.06	&	$\cdots$	\\
5	&	J104210.25+594253.5	&	0.00	&	$\cdots$	\\
6	&	J104526.73+595422.6	&	0.20	&	$\cdots$	\\
7	&	J104556.84+570747.0	&	0.01	&	$\cdots$	\\
8	&	J104625.02+584839.1	&	0.03	&	$\cdots$	\\
9	&	J104633.70+571530.4	&	0.03	&	$\cdots$	\\
10	&	J104840.28+563635.6	&	0.24	&	$\cdots$	\\
11	&	J104857.92+560112.3	&	0.05	&	$\cdots$	\\
12	&	J105000.21+581904.2	&	0.01	&	$\cdots$	\\
13	&	J105106.12+591625.1	&	0.07	&	$\cdots$	\\
14	&	J105518.08+570423.5	&	0.00	&	$\cdots$	\\
15	&	J105604.00+581523.4	&	0.02	&	$\cdots$	\\
16	&	J105959.93+574848.1	&	0.31	&	$\cdots$	\\
17	&	J160015.68+552259.9	&	0.14	&	$\cdots$	\\
18	&	J160128.54+544521.3	&	0.01	&	$\cdots$	\\
19	&	J160341.44+541501.5	&	0.11	&	$\cdots$	\\
20	&	J160523.10+545613.3	&	0.05	&	$\cdots$	\\
21	&	J160630.60+542007.5	&	0.10	&	$\cdots$	\\
22	&	J160908.95+533153.2	&	0.01	&	$\cdots$	\\
23	&	J163031.46+410145.6	&	0.03	&	$\cdots$	\\
24	&	J163135.46+405756.4	&	0.24	&	$\cdots$	\\
25	&	J163143.76+404735.6	&	0.41	&	$\cdots$	\\
26	&	J163352.34+402115.5	&	0.00	&	$\cdots$	\\
27	&	J163502.80+412952.9	&	0.04	&	$\cdots$	\\
28	&	J163854.62+415419.5	&	0.11	&	$\cdots$	\\
29	&	J171126.94+585544.2	&	0.15	&	$\cdots$	\\
30	&	J171334.02+595028.3	&	0.00	&	$\cdots$	\\
31	&	J171736.90+593011.4	&	0.06	&	$\cdots$	\\
32	&	J171748.43+594820.6	&	0.00	&	$\cdots$	\\
33	&	J171818.14+584905.2	&	0.10	&	$\cdots$	\\
34	&	J172104.75+592451.4	&	0.06	&	$\cdots$	\\
35	&	J172414.05+593644.0	&	0.23	&	$\cdots$	\\
\hline					
\end{tabular}
\label{table:tau}
\end{table}

\subsubsection*{Infrared luminosities}

LoBALs span a range of infrared luminosities. Nevertheless, they have median total and starburst infrared luminosities comparable to those of non-LoBALs. The majority of the objects were not detected in the FIR MIPS bands at 70 and 160 $\mu$m. For those, roughly three-fourths of each sample, we have only upper limit estimates of their infrared luminosities. Using the Gehan and the logrank survival statistical tests, we find that the probabilities that the LoBAL and the non-LoBAL samples are drawn from the same distribution are 59\% and 24\%, respectively, considering the slight difference in the median total infrared luminosity, and 79\% and 96\%, respectively, considering the differences in the median starburst infrared luminosity. We conclude that the infrared luminosities of LoBALs are not statistically different from those of non-LoBALs.

\begin{figure}[htpb]
\epsscale{1.0}
\plotone{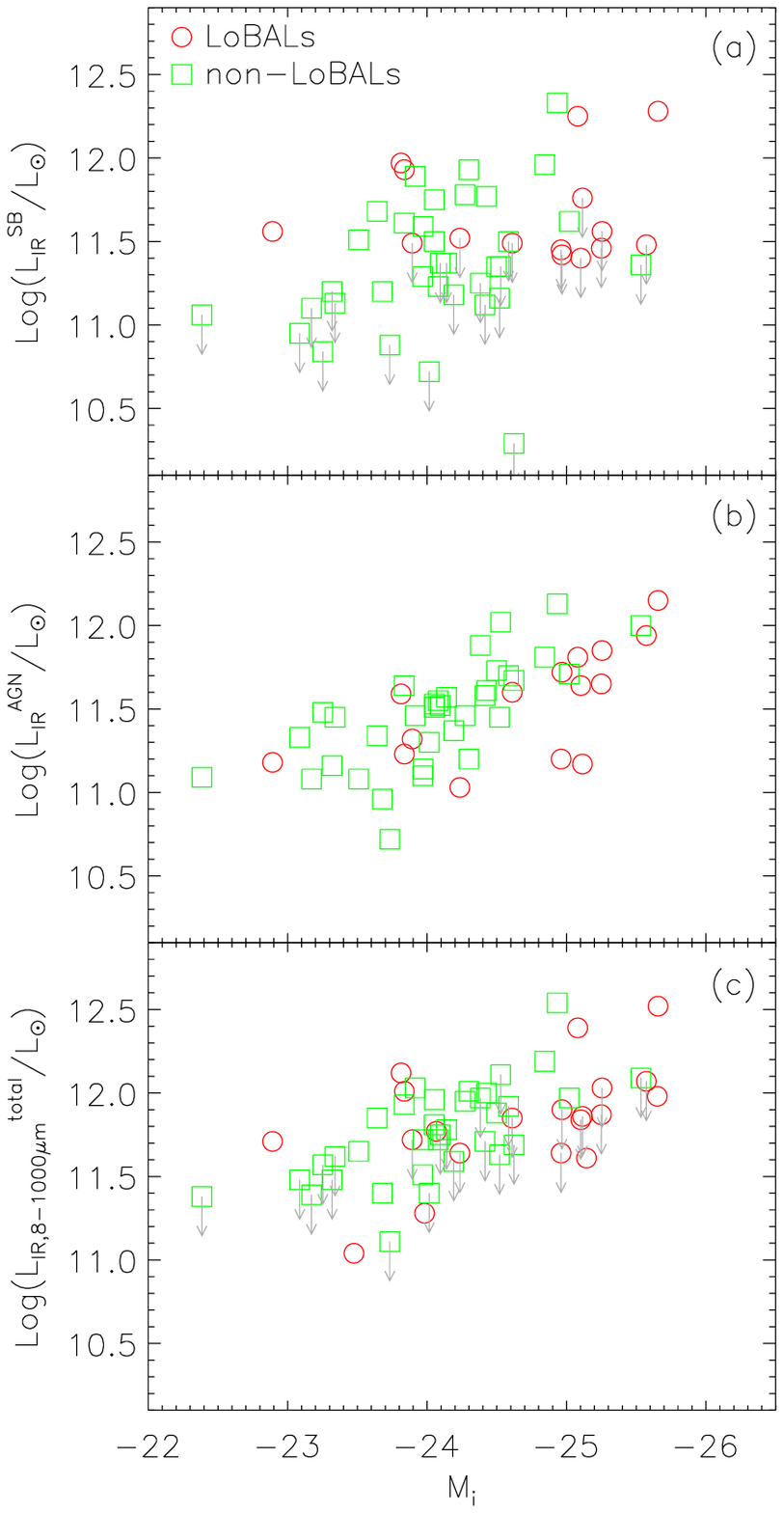}
\caption{\footnotesize Infrared luminosity vs. absolute $M_i$ magnitude corrected for AGN reddening with the extinction estimated from the SED fitting. The LoBALs are plotted with open \textcolor{red}{red circles}; the non-LoBALs are shown with open \textcolor{green}{green squares}. Down arrows indicate upper limits. The FIR limits for the non-LoBAL QSOs are better constrained due to the lower uncertainties in their MIPS images (i.e., non-LoBALs: $\sigma_{70\mu m} \sim$ 1.6 mJy and $\sigma_{160\mu m} \sim$ 3.7 mJy; LoBALs: $\sigma_{70\mu m} \sim$ 6.2 mJy and $\sigma_{160\mu m} \sim$ 8.1 mJy). (a) Starburst infrared luminosity vs. $M_i$. (b) AGN infrared luminosity from 8$-$1000 $\mu$m vs. $M_i$.  The apparent correlation implies comparable IR-to-optical ratios, hence, similar cover fractions in LoBALs and in non-LoBALs. (c) Total infrared luminosity from 8$-$1000 $\mu$m vs. $M_i$. }
\label{fig:mi}
\end{figure}

We find a possible correlation between the extinction-corrected absolute $M_i$ magnitude and the starburst luminosity of the objects with FIR detections (see Fig.~\ref{fig:mi}a). This implies that the star formation rates are higher in the more optically luminous sources. However, since the majority of the measurements are upper limits (22 detections vs.\ 28 upper limits), we are not able to properly quantify the slope or the scatter in the correlation.  Similarly, evidence for a correlation between the optical 5100 $\AA$ luminosity and the 60 $\mu$m IRAS flux was found in low-redshift PG QSOs by \citet{Netzer2007}, who proposed the correlation was mostly due to star formation. We note that the control sample of non-LoBALs is well matched to the absolute $M_i$ magnitudes of the LoBALs (Figure~\ref{fig:mi}) . The range of $M_i$, corrected for Galactic extinction, for the LoBALs is $-22.41 > M_i > -25.55$, with a median of $M_i = -$24.10 (median of the FIR-detections is -23.68). For the control non-LoBALs, the range is $-22.17 > M_i> -25.53$, with a median of $M_i = -$23.96 (median of the FIR-detections is -23.85).

We also find a correlation between the AGN infrared luminosity and the optical i-band luminosity (Fig.~\ref{fig:mi}b), $L_{IR}^{AGN} \sim L_{i}^{\alpha}$ with a slope $\alpha \simeq$ 0.29$\pm$0.05. This relationship holds for both LoBALs and non-LoBALs and covers over three orders of optical magnitude. If the AGN infrared luminosity arises mostly from dusty obscuring material close to the central source, then this correlation implies that LoBALs and non-LoBALs have comparable covering fractions.

\subsubsection*{Star formation}

Star formation contributes a median of 41\% and 48\% of the FIR power in LoBALs and non-LoBALs, respectively, with large variations among individual objects, in agreement with the value found by \citet{Schweitzer2006} for PG QSOs at $z < 0.3$ who claim that star formation contributes at least 30\% of their FIR luminosity. 

 We estimate SFRs from the FIR luminosity solely contributed by the starburst. With the exception of four LoBALs with SFRs $\sim$ 150 $-$ 300 M$_\odot$ yr$^{-1}$, and with the caveat that we only have upper limits for most of the other objects, we find that LoBALs have SFRs $\sim$ 52 M$_\odot$ yr$^{-1}$, slightly higher but comparable to the value we find for non-LoBALs, 38 M$_\odot$ yr$^{-1}$ . However, we note that for the most star-forming galaxies, those with FIR MIPS detections, the median SFR in LoBALs ($\sim$ 161 M$_\odot$ yr$^{-1}$) is higher than the median SFR in non-LoBAL type-1 hosts ($\sim$ 90 M$_\odot$ yr$^{-1}$).

\subsubsection*{LoBALs hitherto and future work}

Overall, our results suggest that LoBALs are very similar to non-LoBALs in terms of their mid-infrared spectral properties and far-infrared luminosities.  This fits the orientation scenario, which would predict comparable levels of star formation for LoBALs and non-LoBALs if both are drawn from the same parent population. Our results, however, cannot rule out an evolutionary paradigm, where LoBALs are rapidly transitioning from a dusty phase marked by high SFRs to a more quiescent phase with SFRs typical of non-BALs. 

The majority of the LoBALs in our sample have SFRs comparable to non-LoBALs, which in the framework of the evolutionary model implies that most of the LoBALs have already passed through the event that quenched the star formation in the galaxy to the levels seen in normal QSO hosts. Our results show that there are large variations among individual LoBALs. Finding that only 20\% of the LoBALs have SFRs 80\% higher than those found in the IR-luminous non-LoBAL implies that the period during which the star formation was quenched was very brief during the short LoBAL transition phase.  If the SFRs in both samples were found to be much higher, but still comparable, a stronger claim could be made that BAL outflows alone are not responsible for quenching the star formation on galactic scale, but that was not observed. 

While the $Spitzer$ observations explore the relation between LoBALs and star formation, they are not sufficient to test whether the LoBAL phenomenon is indeed related to the early stages of QSO activity. Moreover, due to the preponderance of upper limits, our $Spitzer$ results do not allow us to draw strong conclusions about the nature of LoBALs. This study will be complemented by $HST$ imaging and Keck spectroscopy programs of this sample, which will help us study the host galaxy morphologies and the ages of the dominant stellar populations. On one hand, correlating the merger stage with spectral characteristics will give us a deeper insight into the dynamics involved and help us constrain an evolutionary connection between unobscured QSOs and LoBALs. On the other hand, the stellar ages will help us constrain the time scales involved in fueling of the nucleus and the onset (and perhaps quenching) of star formation.

\section{Conclusion}

We investigate the mid- and far-IR properties of a volume-limited sample of 22 low-ionization broad absorption line QSOs within the redshift range $0.5 < z < 0.6$. We model their SEDs from the optical to the far-infrared in an effort to estimate total infrared luminosities, the relative contributions from the starburst and the AGN, starburst luminosities, and star formation rates corrected for the AGN contamination of the FIR emission. We compare this LoBAL sample to a control sample of non-LoBALs, matched by $M_i$ within the redshift $0.45 < z < 0.83$, to examine the possible connection between these two classes of QSOs.  We find that LoBALs are indistinguishable from non-LoBAL type-1 QSO in terms of their MIR spectral properties and FIR luminosities.

\acknowledgments
We are grateful to the anonymous referee for her/his constructive and detailed comments which helped improved both the contents and the presentation of this manuscript. We also thank Christian Leipski for his help in planning the $Spitzer$ observations and Hai Fu for help with IDL. This work is based in part on observations made with the $Spitzer$ Space Telescope (Program ID 50792), which is operated by the Jet Propulsion Laboratory, California Institute of Technology under a contract with NASA. Support for this work was provided by NASA through an award issued by JPL/Caltech.  Additional support was provided by the National Science Foundation, under grant number AST 0507450, and by NASA through a grant from the Space Telescope Science Institute (Program GO-11557), which is operated by the Association of Universities for Research in Astronomy, Incorporated, under NASA contract NAS5-26555. This research has made use of the NASA/IPAC Extragalactic Database (NED) and the NASA/ IPAC Infrared Science Archive, which are operated by the Jet Propulsion Laboratory, California Institute of Technology, under contract with the National Aeronautics and Space Administration. Funding for the Sloan Digital Sky Survey (SDSS) has been provided by the Alfred P. Sloan Foundation, the Participating Institutions, the National Aeronautics and Space Administration, the National Science Foundation, the U.S. Department of Energy, the Japanese Monbukagakusho, and the Max Planck Society. The SDSS Web site is http://www.sdss.org/. The SDSS is managed by the Astrophysical Research Consortium (ARC) for the Participating Institutions. The Participating Institutions are The University of Chicago, Fermilab, the Institute for Advanced Study, the Japan Participation Group, The Johns Hopkins University, Los Alamos National Laboratory, the Max-Planck-Institute for Astronomy (MPIA), the Max-Planck-Institute for Astrophysics (MPA), New Mexico State University, University of Pittsburgh, Princeton University, the United States Naval Observatory, and the University of Washington. This publication makes use of data products from the Two Micron All Sky Survey, which is a joint project of the University of Massachusetts and the Infrared Processing and Analysis Center/California Institute of Technology, funded by the National Aeronautics and Space Administration and the National Science Foundation.

{\it Facilities:} \facility{Spitzer (MIPS)}, \facility{Spitzer (IRS)}

\appendix
\section{APPENDIX: SDSS spectra of the 22 LoBALs}

Figure~\ref{fig:sdds} shows the SDSS spectra of the 22 LoBALs in the sample and details on the \ion{Mg}{2} line to assure the reader that the sample comprises of bona fide LoBALs.  We adopt the more inclusive definition by \citet{Trump2006} requiring the troughs to span a velocity width of at least 1000 km s$^{-1}$ blue-ward of Mg~{\sc II} $\lambda$2800.  Eight of the 22 LoBALs here have blue-shifted \ion{Mg}{2} absorption line widths between 1000 and 2000 km s$^{-1}$, which would classify them as mini-BALs according to the traditional definition by \citet{Weymann1991} requiring widths greater than 2000 km s$^{-1}$.

\begin{figure}[htpb]
\epsscale{1.0}
\plotone{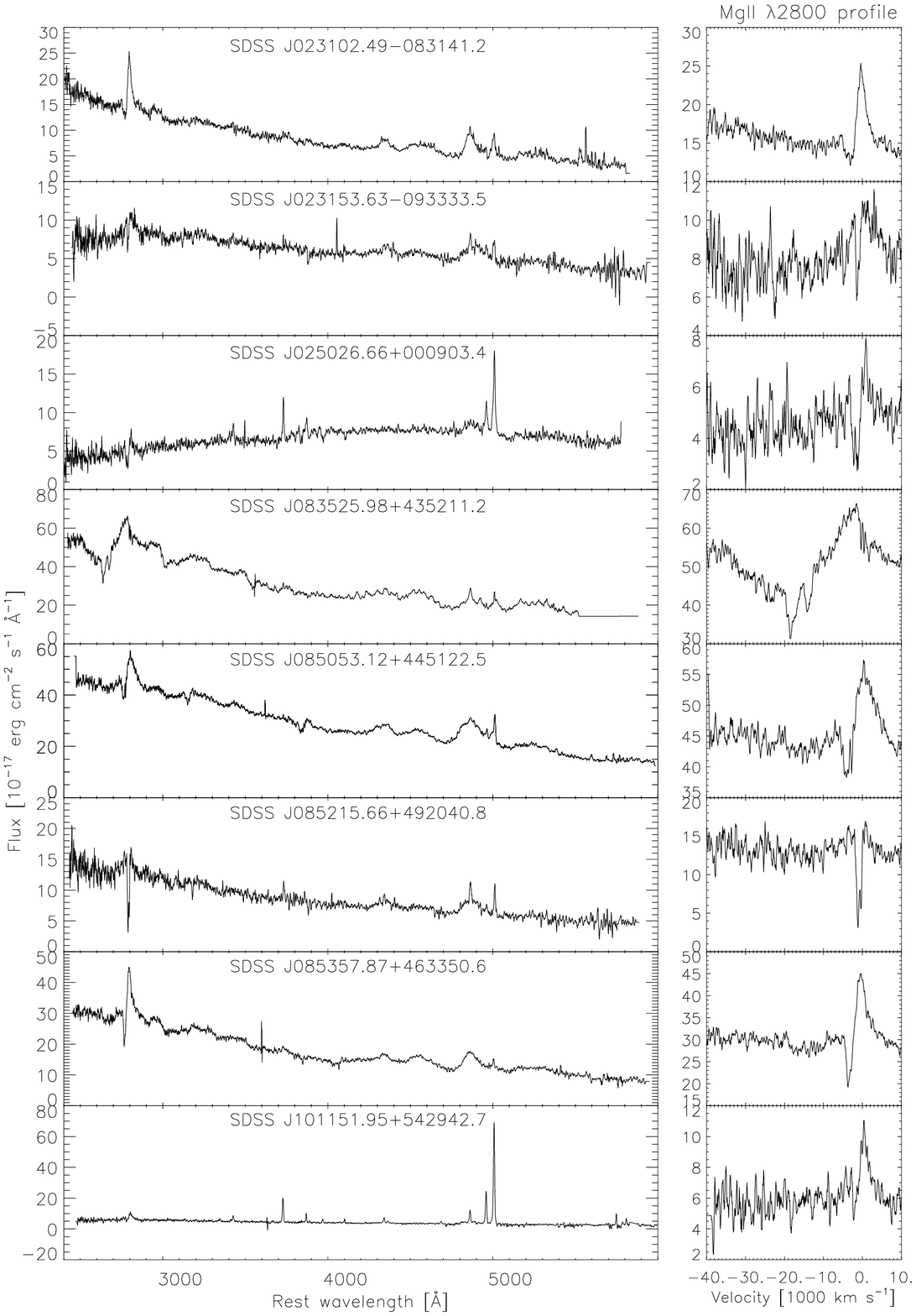}
\caption{\footnotesize Left: SDSS spectra of the 22 LoBALs. Right: Velocity profile of the \ion{Mg}{2} $\lambda$2800 line.}
\label{fig:sdds}
\end{figure}

\addtocounter{figure}{-1}
\begin{figure}[htpb]
\epsscale{1.}
\plotone{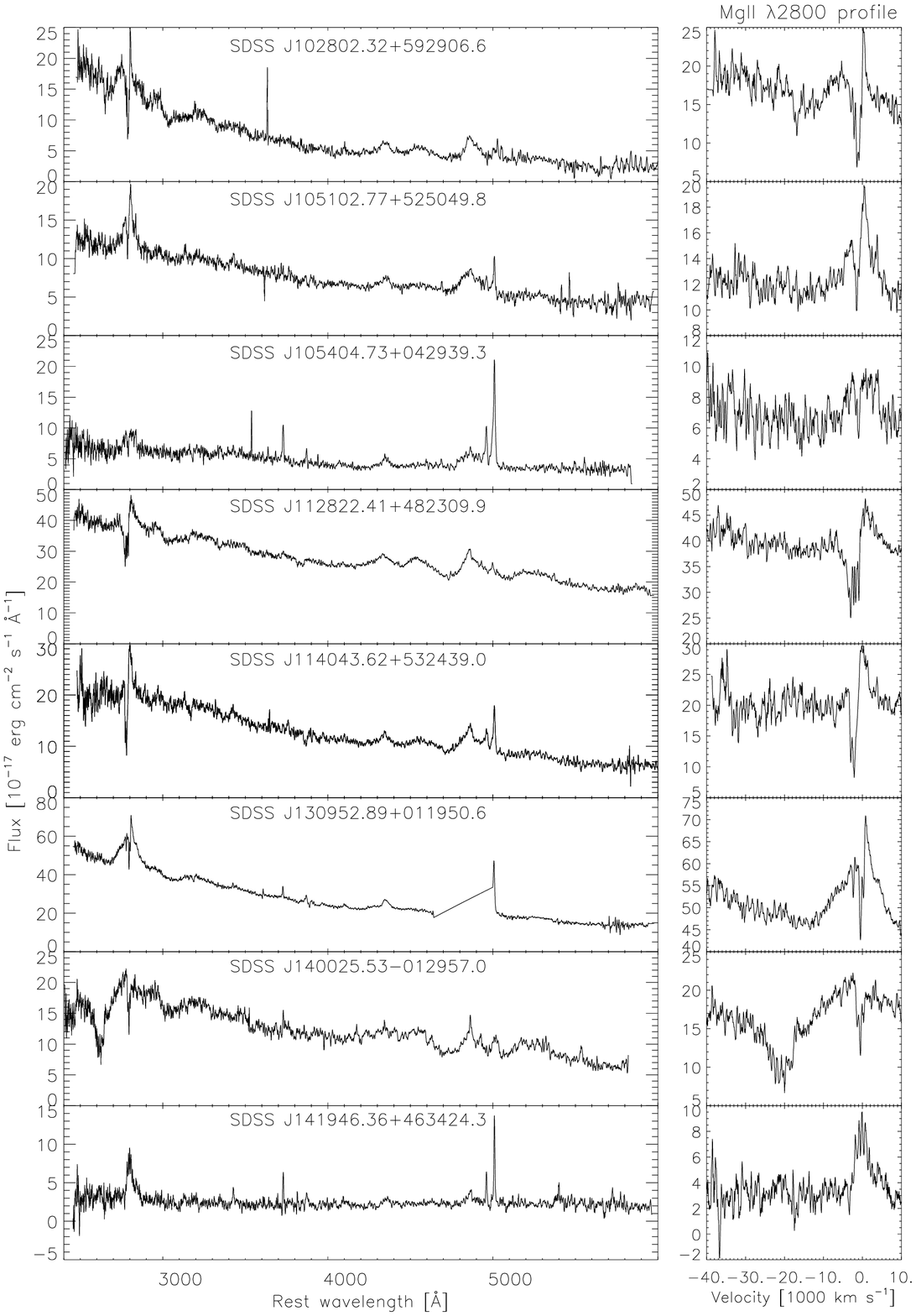}
\caption{Continued.}
\end{figure}

\addtocounter{figure}{-1}
\begin{figure}[htpb]
\epsscale{1.0}
\plotone{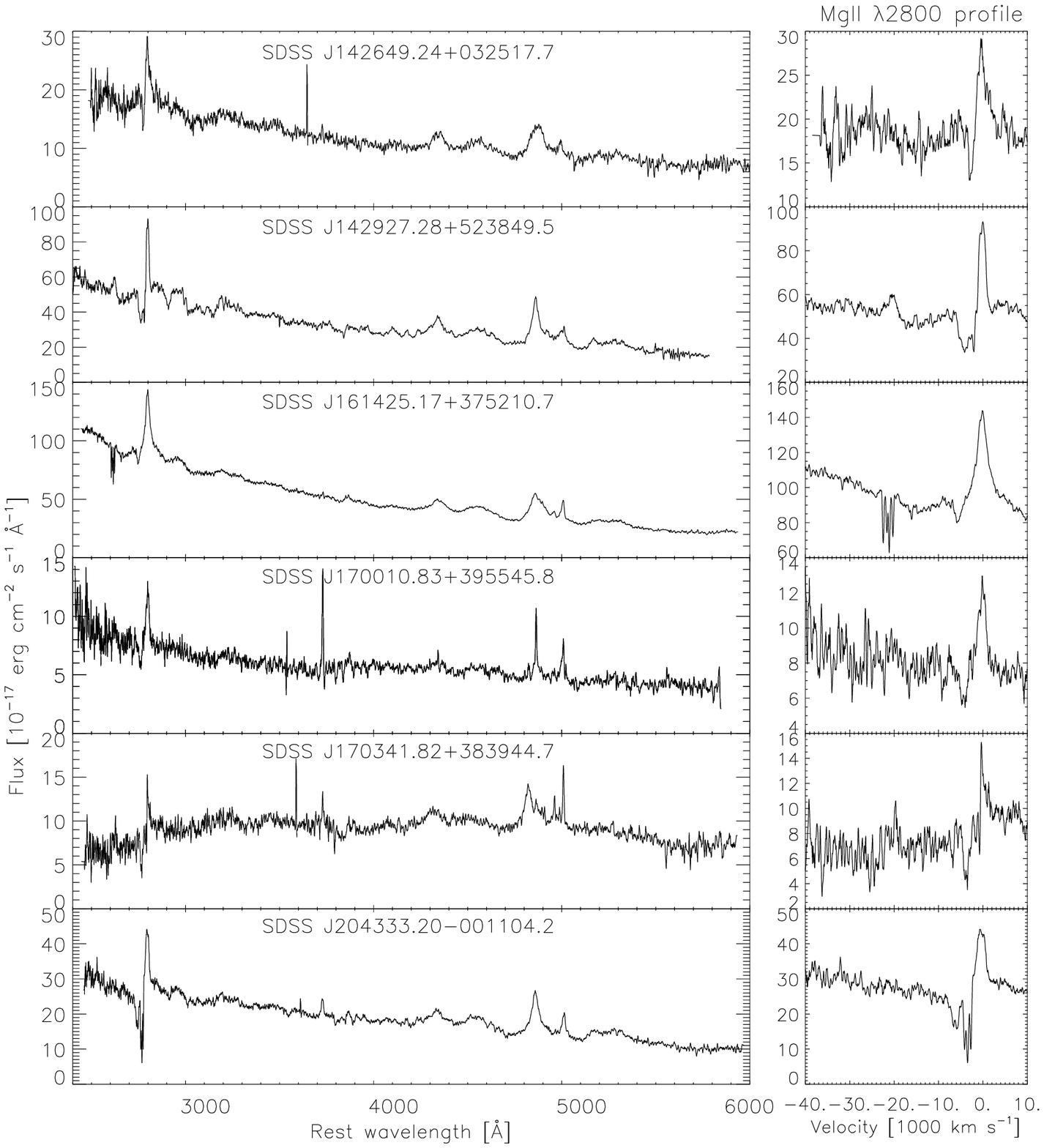}
\caption{Continued.}
\end{figure}

\bibliographystyle{apj}
\bibliography{REF}

\end{document}